\documentclass[showpacs,twocolumn]{revtex4}

\usepackage{graphicx}
\usepackage{float}
\usepackage{dcolumn}
\usepackage{amsmath}
\usepackage{multirow}
\usepackage{amsfonts}
\usepackage{multirow}
\usepackage{booktabs}
\usepackage{makecell}
\usepackage{color}
\usepackage[dvipdfm,colorlinks,
            linkcolor=blue,
            anchorcolor=blue,
            citecolor=blue]{hyperref}

\makeatletter
\def\btt#1{\texttt{\@backslashchar#1}}
\DeclareRobustCommand\bblash{\btt{\@backslashchar}} \makeatother

\begin{document}
\title{Critical behavior of AdS black holes surrounded by dark fluid with Chaplygin-like equation of state}
\author{Xiang-Qian Li $^{a}$} \email{lixiangqian@tyut.edu.cn}
\author{Hao-Peng Yan $^{a}$} \email{yanhaopeng@tyut.edu.cn}
\author{Li-Li Xing $^{a}$} \email{xinglili@tyut.edu.cn}
\author{Shi-Wei Zhou $^{b}$} \email{zhousw783@163.com}
\affiliation{$^{a}$ College of Physics, Taiyuan University of Technology, Taiyuan 030024, China}
\affiliation{$^{b}$ Physics and Information Engineering Institute, Shanxi Normal University, Taiyuan 030031, China}
\date{\today}

\begin{abstract}
Supposing the existence of Dark Fluid with a Chaplygin-like equation of state $p=-B/\rho$ (CDF) as a cosmic background, we obtain a static spherically-symmetric black hole (BH) solution to the Einstein gravitational equations. We study the $P-V$ critical behavior of AdS BH surrounded by the CDF in the extended phase space where the cosmological constant appears as pressure, and our results show the existence of the Van der Waals like small/large BH phase transition. Also, it is found that such a BH displays a first-order low/high-$\Phi$ BH phase transition and admits the same criticality with van der Waals liquid/gas system in the non-extended phase space, where the normalization factor $q$ is considered as a thermodynamic variable, while the cosmological constant being fixed. In both $P-V$ and the newly proposed $q-\Phi$ phase spaces, we calculate the BH equations of state and then numerically study the corresponding critical quantities. Moreover, the critical exponents are derived and the results show the universal class of the scaling behavior of thermodynamic quantities near criticality. Finally, we study the shadow thermodynamics of AdS BHs surrounded by the CDF. We find that, there exists a positive correlation between the shadow radius and the event horizon radius in our case. By analyzing the temperature and heat capacity curves under the shadow context, we discover that the shadow radius can replace the event horizon radius to demonstrate the BH phase transition process, and the changes of the shadow radius can serve as order parameters for the small/large BH phase transition, indicating that the shadow radius could give us a glimpse into the BH phase structure from the observational point of view.
\end{abstract}

\maketitle

\section{Introduction}

Matter content of the Universe is still an unsolved problem in the framework of modern cosmology. The latest release of 2018 Planck full-sky maps about the CMB anisotropies \cite{Planck2018} illustrates that baryon matter component is no more than $5\%$ for total energy density. In contrast, the invisible dark components, including dark energy and dark matter, are about $95\%$ energy density in the Universe. The dominance of the dark sector over the Universe makes the study of BHs surrounded by these mysterious field well-deserved. Quintessence is a possible candidate for dark energy, which is characterized by the linear equation of state $p=\omega\rho$ with $\omega$ a constant in the range of $-1<\omega<-1/3$. Significant attention has been devoted to discussion of static spherically-symmetric BH solutions surrounded by quintessence matter and their properties \cite{KiselevCQG2003,MaCPL2007,FernandoGRG2012,FengPLA2014,MalakolkalamiASS2015,HussainGRG2015,Ghosh:2016ddh,Ghosh:2017cuq,Toledo:2018pfy}. Besides the quintessence matter, many authors have found exact BH solutions with some other sources. The BH solutions with the existence of an Abelian gauge field for which the density is given by a power of Maxwell Lagrangian has been introduced in Ref.~\cite{Hassaine:2007py}. The solutions of general relativity coupled with nonlinear electrodynamics, consisting of Einstein-Born-Infeld solutions in asymptotically flat space~\cite{Hoffmann:1935ty} and (A)dS spaces~\cite{Dey:2004yt,Cai:2004eh}, have been studied. The Yang-Mills theory has been coupled to a gravitating system and the resulted various BH solutions using such a coupled theory has been studied in Refs.~\cite{Bizon:1990sr,Kuenzle:1990is}. Exact BH solutions with string cloud backgrounds in general relativity have been found in Ref.~\cite{Letelier:1979ej}. Study combining the dark matter in the quintessence form with the cloud of strings was recently proposed for charged AdS BH~\cite{Toledo:2019amt}.

With regard to the Universal dark sector, there exists another possibility that the unknown energy component is a unified dark fluid which mixes dark matter and dark energy. Among the proposed unified dark fluid models, the Chaplygin gas \cite{Kamenshchik:2001cp} and its generalized models \cite{Bilic:2001cg,Bento:2002ps} have been widely studied in order to explain the accelerating Universe~\cite{Carturan:2002si,Amendola:2003bz,Bean:2003ae}.
Therefore, it is interesting to adopt (generalized) Chaplygin gas as a matter source to construct BH solutions. Recently, we considered a model that charged static spherically-symmetric BH is surrounded by CDF in the framework of Lovelock gravity, the analytical BH solution was deduced and the related thermodynamic quantities were calculated~\cite{Lovelockdf2019}. Also, we extended our study to the modified Chaplygin gas (MCG) case, special attention has been paid to the thermodynamical stability of the MCG-surrounded BHs in Einstein-Gauss-Bonnet gravity~\cite{EGBMCG2022} and Lovelock theory of gravity~\cite{Li:2022csn}. Here we should note that, though the Chaplygin gas model is usually introduced in cosmological studies to account for the cosmological evolution, its equation of state can be deduced naturally in string theory. The equation of state $p=-\frac{B}{\rho}$ can be obtained from the Nambu-Goto action for a $d$-brane moving in a $(d+2)$-dimensional spacetime in light-cone parametrization~\cite{Ogawa:2000gj,Bordemann:1993ep}, and the Chaplygin gas is verified to admit a supersymmetric generalization~\cite{Jackiw:2000cc}. Besides, the Chaplygin gas manifests itself as the effect of the immersion of our four-dimensional world into some multi-dimensional bulk~\cite{Gorini:2004by}. The Chaplygin equation of state also arises in connection with the Randall-Sundrum model~\cite{Randall:1999vf}. Thus a CDF could be a naturally existing substance, more than a phenomenological model designed for cosmological applications. In this paper, we focus on the phase transitions and critical behaviors of the static spherically-symmetric AdS BHs surrounded by CDF in the Einstein's theory of gravity.

This paper is organized as follows. In Section \ref{section2}, for a CDF, we deduce its stress-energy tensor, with the help of which we obtain a static spherically-symmetric solution to the Einstein field equations. Further we analyze the critical thermodynamical behaviors of the newly derived BH solution in $P-V$ phase space in Section \ref{section3}. Also, we analyze the $q-\Phi$ criticality of AdS BHs surrounded by CDF in Section \ref{section4}. In Section \ref{section5}, we examine the phase transitions of CDF-surrounded AdS BHs by using shadow analysis. Section \ref{section6} gives the conclusion.

We use units which fix the speed of light and the gravitational constant via $8\pi G = c = 1$, and use the metric signature ($-,\;+,\;+,\;+$).

\section{BH solutions in the background of CDF}
\label{section2}
With regard to the case that a static spherically-symmetric BH has an atmosphere composed of field with explicit Lagrangian, it is convenient to study the interplay between the BH spacetime and field by jointly solving the gravitational field equations and equation of motion of the concerned field~\cite{Hassaine:2007py,Hoffmann:1935ty,Dey:2004yt,Cai:2004eh,Bizon:1990sr,Kuenzle:1990is,Letelier:1979ej,Toledo:2019amt}. However, if the nature of atmosphere matter is unclear, e.g., quintessence dark energy~\cite{KiselevCQG2003} and (generalized) Chaplygin gas~\cite{Lovelockdf2019,EGBMCG2022,Li:2022csn}, it is essential to consider how to study the interaction between the curvature in spacetime and the matter with only knowing the equation of state of the matter fluid.

\subsection{Cosmological dark fluid with equation of state $p=-B/\rho$}
Considering a static spherically-symmetric spacetime, we adopt the following form for the metric
\begin{equation}
ds^2=-f(r)dt^2+\frac{1}{g(r)} dr^2+r^2 d\Omega^2,\label{dsf}
\end{equation}
where $f(r)$ and $g(r)$ are general functions depending on the radial coordinate $r$ and $d\Omega^2=d\theta^2+{\rm sin}^2\theta d\phi^2$ stands for the standard element on $S^2$.

The stress-energy tensor for a perfect fluid is
\begin{equation}
T_{\mu\nu}= (\rho+p) u_\mu u_\nu +pg_{\mu\nu}, \label{Stress Energy Tensor}
\end{equation}
where $\rho$ and $p$ are energy density and isotropic pressure, respectively, as measured by an observer moving with the fluid, and $u_\mu$ is its four-velocity. In this regard, the static spherically-symmetric solutions of Einstein's equations for perfect fluid source (dust, radiation, dark energy or phantom energy) with equation of state $p=\omega p$ ($\omega$ is a constant) have been studied by Semiz~\cite{Semiz:2008ny}. There is another scenario with respect to the pressures of the fluid, the cosmological fluid surrounding a BH could be anisotropic because of the gravitational attraction near the central body. Along this line, Kiselev~\cite{KiselevCQG2003} obtained a BH spacetime, which soon became a remarkably popular toy model, by treating the ambient quintessence mattar as anisotropic fluid. For the CDF, even its generating mechanism, from a field theoretical point of view, is not clearly identified, there are several likely candidates in string theory context, which we mentioned in the introduction part, and in phenomenologically cosmological studies. Confronting with the cosmological evolution, the CDF is usually modeled by introducing a scalar field $\varphi$ and a self-interacting potential $U(\varphi)$, with the Lagrangian $L_{\varphi}=-\frac{1}{2}\partial_{\mu}\varphi\partial^{\mu}\varphi-U(\varphi)$~\cite{Gorini:2004by,Debnath:2004cd,Mak:2005iq}. Also, the fluid representation of the CDF can be recast under the form of a tachyonic field $T$ given by a Born-Infeld type Lagrangian $L_{T}=-V(T)[1+\partial_{\mu}T\partial^{\mu}T]^{1/2}$ with $V(T)$ one arbitrary function~\cite{Sharif:2014yga,MagalhaesBatista:2009cus}. Else, it is found that the (modified) Chaplygin gas model can be reconstructed by $k$-essencee (kinetic quintessence)~\cite{Chimento:2003ta,Sharif:2012cu} and $f$-essence (fermionic $k$-essence)~\cite{Tsyba:2011ss}. Considring the presence of kinetic terms in these viable theories and the fact that the essential field of CDF depends only on the raidal coordinate in static sphrical symmetry, the radial pressure of CDF should be different from the tangential one. In this work, we suppose the CDF to be anisotropic and its stress-energy tensor can be written in a covariant form~\cite{Raposo:2018rjn} as
\begin{equation}
T_{\mu\nu}= \rho u_\mu u_\nu +p_r k_\mu k_\nu +p_t \Pi_{\mu\nu}, \label{Stress Energy Tensor}
\end{equation}
where $p_r$ and $p_t$ are respectively the radial and the tangential pressure, $u_\mathrm{\mu}$ is the fluid four-velocity and $k_\mathrm{\mu}$ is a unit spacelike vector orthogonal to $u_\mathrm{\mu}$, with $u_\mathrm{\mu}$ and $k_\mathrm{\mu}$ satisfying \mbox{ $u_\mathrm{\mu} u^{\mathrm{\mu}}=-1$}, \mbox{ $k_\mathrm{\mu}k^\mathrm{\mu}=1$} and $u^{\mathrm{\mu}}k_\mathrm{\mu}=0$.
$\Pi_\mathrm{\mu\nu} = g_\mathrm{\mu\nu}+u_\mathrm{\mu}u_\mathrm{\nu}-k_\mathrm{\mu}k_\mathrm{\nu}$
is a projection tensor onto a two surface orthogonal to $u^\mathrm{\mu}$ and $k^\mathrm{\mu}$. Working in the comoving frame of the fluid, one obtains that $u_\mathrm{\mu} = (-\sqrt{f},0,0,0)$ and $k_\mathrm{\mu} = (0,1/\sqrt{g},0,0)$. In this way, the stress-energy tensor in Eq.~(\ref{Stress Energy Tensor}) can be reexpressed as
\begin{equation}
T_{\mu}{}^{\nu}=-(\rho+p_t)\delta_{\mu}{}^0\delta^{\nu}{}_0 +p_t \delta_{\mu}{}^{\nu} +(p_r-p_t)\delta_{\mu}{}^1\delta^{\nu}{}_1, \label{Stress Energy Tensor2}
\end{equation}
where the difference between radial and tangential pressures $p_r-p_t$ is known as anisotropic factor. In the limit of $p_r=p_t$, the stress-energy tensor reduces to the standard isotropic form.

Now we consider a matter fluid across an event horizon described by the stress-energy form in Eq.~(\ref{Stress Energy Tensor2}). Inside the horizon, since $g_{tt}>0$ and $g_{rr}<0$, the coordinate $r$ plays the role of time, then the energy density yields $-{{T}_r}^r=-p_r$, and the pressure along the spatial $t$ direction should be ${{T}_t}^t=-\rho$. Considering this exchange of roles, the energy density and pressure are continuous across the horizon only if $p_r=-\rho$. In the case of $p_r\neq-\rho$ and $\rho(r_h)\neq0$, the pressure must be discontinuous at the horizon $r_h$ and thus the solution be dynamical. Here, we require $p_r=-\rho$ in this study, so that the CDF stays static and the energy density is continuous across the horizon, which places a constraint on the solution. In fact, anisotropy and $p_r=-\rho$ are also the case for static Reissner-Nordstr\"{o}m solution, power of Maxwell solution~\cite{Hassaine:2007py}, Einstein-Born-Infeld solutions~\cite{Hoffmann:1935ty,Dey:2004yt,Cai:2004eh}, Yang-Mills solution~\cite{Bizon:1990sr,Kuenzle:1990is} and string cloud solution~\cite{Letelier:1979ej}. For cosmological fluid with a general form of equation of state $p=p(\rho)$, even it shows anisotropy in the gravitational field generated by a BH, its equation of state should appear as $p=p(\rho)$ at cosmological scale, thus one can constrain the tangential pressure $p_t$ by taking isotropic average over the angles and requiring $\langle {T}_i{}^j\rangle=p(\rho)\delta_i{}^j$, that is to say
\begin{equation}
p_t+\frac{1}{3}(p_r-p_t)=p(\rho),\label{angles_average}
\end{equation}
where the relation $\langle \delta_i{}^1\delta^j{}_1\rangle=\frac{1}{3}$ has been used. For the quintessence matter with equation of state $p =\omega\rho$ ($-1<\omega<-1/3$), the tangential pressure can be deduced from Eq.~(\ref{angles_average}) as $p_t=\frac{1}{2}(1+3\omega)\rho$, compatible with the radial pressure $p_r=-\rho$, which is just the result obtained by Kiselev~\cite{KiselevCQG2003}.

In our case, the CDF has a non-linear equation of state $p=-\frac{B}{\rho}$, where $B$ is a positive constant. For $p_r=-\rho$, the tangential pressure yields $p_t=\frac{1}{2}\rho-\frac{3B}{2\rho}$. Thus the stress-energy tensor of the CDF can be expressed as
\begin{eqnarray}
&{{T}_t}^t={{T}_r}^r=-\rho, \label{Ttr}~\\
&{{T}_{\theta}}^{\theta}={{T}_{\phi}}^{\phi}=\frac{1}{2}\rho-\frac{3B}{2\rho}.~\label{Tangular}
\end{eqnarray}
As we shall see later, the anisotropy of the CDF fades away and the equation of state yields $p=-B/\rho$ at cosmological scale.

\subsection{Exact static spherically-symmetric solution}
Since we demand ${{T}_t}^t={{T}_r}^r$, without any lose of generality, the relation between the metric components $g(r)=f(r)$ can be performed by an appropriate rescaling of time. Then the components of the Einstein tensor are given by
\begin{eqnarray}
&{{G}_t}^t={{G}_r}^r=\frac{1}{r^2}(f+rf'-1),~\label{Gtr}\\
&{{G}_{\theta}}^{\theta}={{G}_{\phi}}^{\phi}=\frac{1}{2r}(2f'+rf''). \label{Gthetafai}
\end{eqnarray}

Combining Eqs.~(\ref{Ttr})-(\ref{Tangular}) and~(\ref{Gtr})-(\ref{Gthetafai}), one obtains the gravitational equations:
\begin{eqnarray}
&\frac{1}{r^2}(f+rf'-1)+\Lambda=-\rho,~\label{graviequationsrt}\\
&\frac{1}{2r}(2f'+rf'')+\Lambda=\frac{1}{2}\rho-\frac{3B}{2\rho},~\label{graviequationsthetafai}
\end{eqnarray}
here we consider the presence of cosmological constant. Thus, we have two unknown functions $f(r)$ and $\rho(r)$, which can be determined analytically by the above two differential equations. Now, by solving the set of differential equations (\ref{graviequationsrt})-(\ref{graviequationsthetafai}), one first easily obtains the solution for the energy density of CDF
\begin{equation}
\rho(r)=\sqrt{B+\frac{q^2}{r^6}}, \label{CDFenergydensity}
\end{equation}
where $q>0$ is a normalization factor that indicates the intensity of the CDF. Besides, Eq.~(\ref{CDFenergydensity}) is also the direct result of the conservation law for the stress-energy tensor $\nabla_{\nu}T^{\mu\nu}=0$. We see that for small radial coordinate (i.e. $r^6\ll q^2/B$), the CDF energy density is approximated by
\begin{equation}
\rho(r)\approx\frac{q}{r^3}, \label{CDFenergydensitysmallr}
\end{equation}
indicating that the CDF behaves like a matter content whose energy density vary with $r^{-3}$. At large radial coordinate (i.e. $r^6\gg q^2/B$), it follows that
\begin{equation}
\rho(r)\approx \sqrt{B}, \label{CDFenergydensitylarger}
\end{equation}
meaning that the CDF acts like a positive cosmological constant at large scale ragime. One can also observe that $p_r\rightarrow-\sqrt{B}$ and $p_{\theta,\phi}\rightarrow-\sqrt{B}$ when $r\rightarrow\infty$, indicating that the CDF appears to be isotropic and its equation of state recovers $p=-B/\rho$ at cosmological scale. We note that, for a cosmological fluid with general equation of state $p=p(\rho)$, whose radial pressure satisfying $p_r=-\rho$ when surrounding a central BH, it always tends to be isotropic at cosmological scale. As shown in Table~\ref{CDFvsQM}, the anisotropic factor $p_r-p_t$ for both the CDF and quintessence matter in Kiselev's solution~\cite{KiselevCQG2003} reduce to zero at infinity.

\begin{table*}[htbp]
\tabcolsep 0pt
\caption{CDF and quintessence matter in the spacetime of Einstein(-AdS) BH.}
\vspace*{-12pt}
\begin{center}
\def\temptablewidth{0.8\textwidth}
{\rule{\temptablewidth}{1pt}}
\renewcommand\arraystretch{1.2}
\begin{tabular*}{\temptablewidth}{@{\extracolsep{\fill}}cccccc}
    Anisotropic fluid & EoS & $p_r$ & $p_t$ & $\rho$ & Asymptotic behavior at infinity   \\   \hline
    \makecell[c]{Quintessence\\
     matter~\cite{KiselevCQG2003} } & \makecell[c]{$p=\omega\rho$\\
     $(-1<\omega<-1/3)$} & $-\rho$ & $\frac{1}{2}(1+3\omega)\rho$& $-\frac{a}{2}\frac{3\omega}{r^{3(\omega+1)}}$ & $\rho\rightarrow0$, $p_r\rightarrow0$, $p_t\rightarrow0$  \\
    CDF & \makecell[c]{$p=-\frac{B}{\rho}$\\
     $(B>0)$}& $-\rho$ & $\frac{1}{2}\rho-\frac{3B}{2\rho}$ & $\sqrt{B+\frac{q^2}{r^6}}$ & $\rho\rightarrow\sqrt{B}$,  $p_r\rightarrow-\sqrt{B}$, $p_t\rightarrow-\sqrt{B}$   \\
\end{tabular*}
{\rule{\temptablewidth}{1pt}}
\end{center}
\label{CDFvsQM}
\end{table*}

\begin{figure}[htb]
\includegraphics[width=0.9\linewidth]{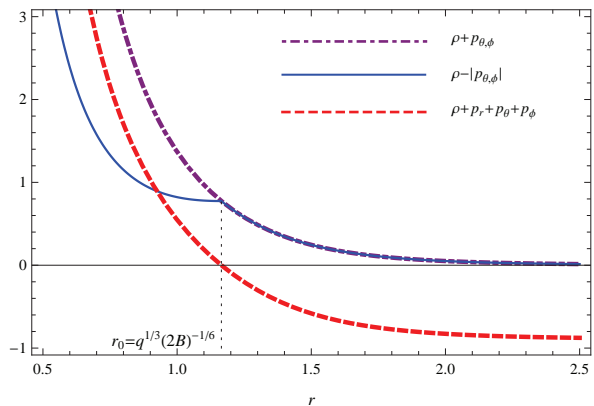}
\caption{\label{energycondition} The variation of $\rho+p_{\theta,\phi}$, $\rho-|p_{\theta,\phi}|$ and $\rho+p_r+p_{\theta}+p_{\phi}$ versus $r$ for the CDF taking $q=1.0$ and $B=0.2$.}
\end{figure}

Energy conditions are very useful tools to discuss cosmological geometry~\cite{Visser:1999de} and BH spacetimes~\cite{Zaslavskii:2010qz,Neves:2014aba} in both general relativity~\cite{Kontou:2020bta} and modified gravity~\cite{Capozziello:2013vna}. The standard energy conditions include null energy condition (NEC), weak energy condition (WEC), strong energy condition (SEC), and dominant energy condition (DEC), given as
\begin{eqnarray}
\text{NEC}:&~&\rho+p_i\geq0~(i=r,\theta,\phi); \nonumber \\
\text{WEC}:&~&\rho\geq0~~~\&~~~\rho+p_i\geq0~(i=r,\theta,\phi); \nonumber \\
\text{SEC}:&~&\rho+{\sum}_ip_i\geq0~~~\&~~~\rho+p_i\geq0~(i=r,\theta,\phi); \nonumber \\
\text{DEC}:&~&\rho\geq0~~~\&~~~|p_i|\leq\rho~(i=r,\theta,\phi). ~~~\label{Econdition}
\end{eqnarray}
Relevant quantities are deduced as
\begin{eqnarray}
& \rho+p_r=0,~~~\rho+p_{\theta,\phi}=\frac{3}{2}\left(\rho-\frac{B}{\rho}\right), \nonumber \\
& \rho+p_r+p_{\theta}+p_{\phi}=\rho-\frac{3B}{\rho}, \nonumber \\
& \rho-|p_r|=0,~~~\rho-|p_{\theta,\phi}|=\rho-\left|\frac{1}{2}\rho-\frac{3B}{2\rho}\right|. ~~~\label{Econditionquantities}
\end{eqnarray}
To examine the energy conditions of the CDF in our case, we plot the unspecified quantities in Eq.~(\ref{Econditionquantities}) with respect to the radial coordinate $r$ in Fig.~\ref{energycondition}.
It can be observed that, $\rho+p_{\theta,\phi}$ and $\rho-|p_{\theta,\phi}|$ remain positive everywhere, while the sign of $\rho+p_r+p_{\theta}+p_{\phi}$ is converted from positivity to negativity at $r_0=q^{1/3}\left(2B\right)^{-1/6}$, which is exactly the transition point for the sign of $p_{\theta,\phi}$, i.e. the point at which the tangential pressure exhibits a transition from being attractive into being repulsive. We conclude that the CDF well satisfies the NEC, WEC and DEC, however it violates the SEC. This is also the case for the quintessence matter~\cite{KiselevCQG2003}. In fact, a violation of the SEC is equivalent to a violation of the attractive character of gravity~\cite{Ansoldi:2008jw}, as shown by the dark energy which accelerates the expansion of the Universe in cosmological studies~\cite{Curiel:2014zba}, as well as the matter content in the background of a regular BH whose singularity replaced by a de Sitter core~\cite{Conboy:2005nx}.

\begin{figure*}[htb]
\begin{tabular}{ c c c }
\includegraphics[width=0.43\linewidth]{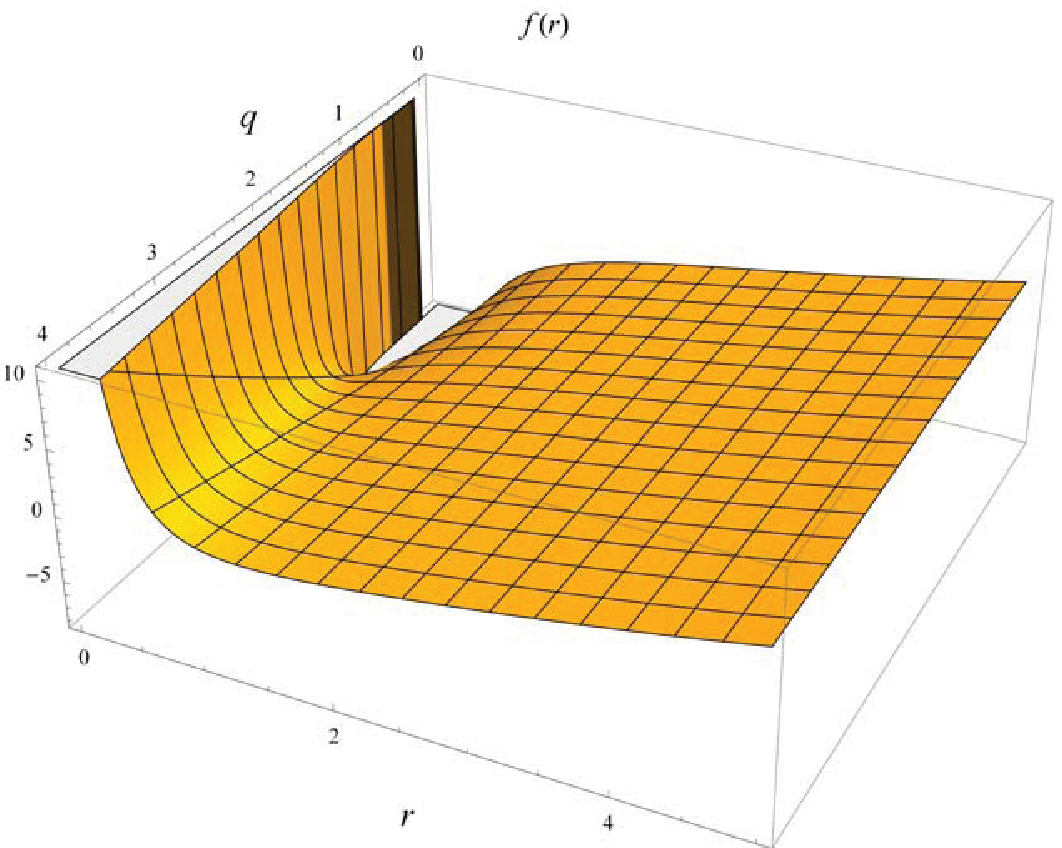}
\hspace{.10in}
\includegraphics[width=0.43\linewidth]{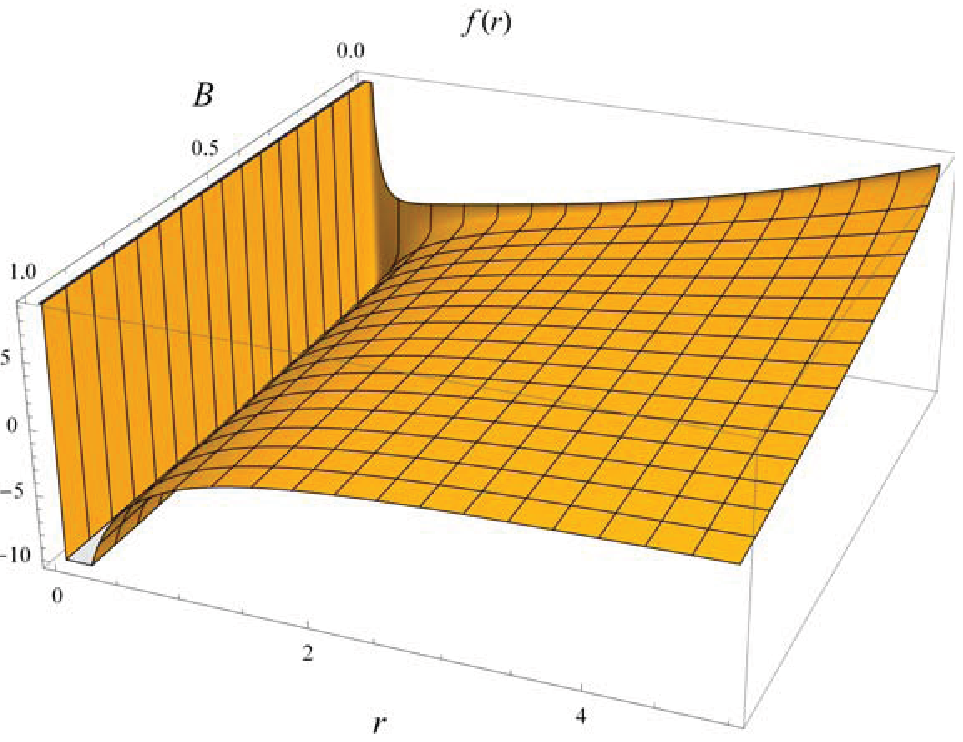}
\end{tabular}
\caption{\label{frwithqB} Left panel: The behavior of $f(r)$ with varying $q$ and fixing $B=0.2$. Right panel: The behavior of $f(r)$ with varying $B$ and fixing $q=1.0$. For both panels, we have set $M=2.0$ and $\Lambda=-1.0$.}
\end{figure*}

Substituting Eq.~(\ref{CDFenergydensity}) into Eq.~(\ref{graviequationsrt}), we obtain the analytical solution for $f(r)$
\begin{equation}
f(r)=1-\frac{2M}{r}-\frac{r^2}{3}\sqrt{B+\frac{q^2}{r^6}}+\frac{q}{3r}{\rm ArcSinh}\frac{q}{\sqrt{B}r^3}-\frac{ r^2}{3}\Lambda, \label{frsolution}
\end{equation}
where $M$ denotes the mass of the BH, here we consider the BH as a point mass BH, thus $M$ arises as a constant. One can examine that this solution for $f(r)$ satisfies Eq.~(\ref{graviequationsthetafai}). To study the asymptotic behavior of $f(r)$, we take $r\rightarrow\infty$ and find that
\begin{equation}
f(r)\rightarrow 1-\frac{r^2}{3}\left(\sqrt{B}+\Lambda\right),
\end{equation}
which reveals that, the asymptotic behavior of the solution is determined by both the cosmological constant $\Lambda$ and the CDF parameter $B$. In this work, we concern the AdS BH, thus we constrain $\Lambda<-\sqrt{B}$. The dependencies of the metric function $f(r)$ on the parameters $q$ and $B$ are depicted in Fig.~\ref{frwithqB}.

\subsection{Thermodynamics of AdS BH surrounded by CDF}
\label{section3}
\begin{figure}[htb]
\includegraphics[angle=-90,width=0.9\linewidth]{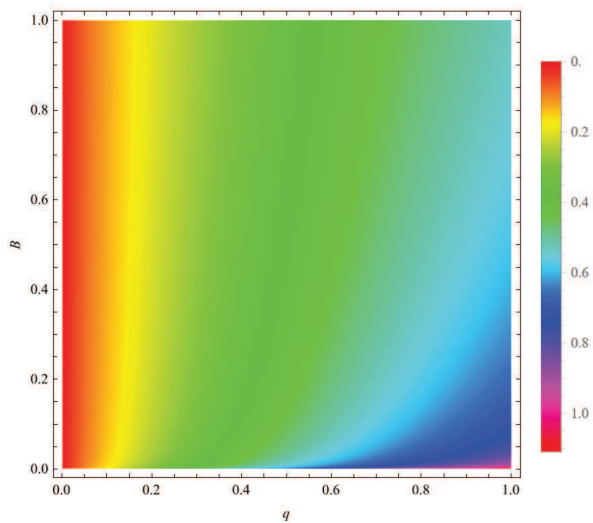}
\caption{\label{extremalBHmass} The extremal BH mass with varying $q$ and $B$. Here we set $\Lambda=-1$. For a physical BH with given $q$, $B$ and $\Lambda$, its mass should be equal to at least the extremal BH mass.}
\end{figure}

Solving the equation $f(r_h)=0$, one can obtain the event horizon radius $r_h$, with which the mass of the BH can be expressed as
\begin{equation}
M=\frac{r_h}{2}-\frac{r_h^3}{6}\Lambda-\frac{r_h^3}{6}\sqrt{B+\frac{q^2}{r_h^6}}+\frac{q}{6}{\rm ArcSinh}\frac{q}{\sqrt{B}r_h^3}.\label{BHMass}\\
\end{equation}
The Hawking temperature can be derived as
\begin{equation}
T=\frac{f'(r_h)}{4\pi}=\frac{1}{4\pi}\left(\frac{1}{r_h}-r_h\sqrt{B+\frac{q^2}{r_h^6}}-r_h\Lambda\right).\label{BHtemperature}
\end{equation}
It can be confirmed through numerical methods that, with fixed values of $q$, $B$, and $\Lambda$, the Hawking temperature has only one zero crossing, which corresponds to the extremal BH case. Since the physical temperature should be non-negative, the mass of a BH must be equal to or greater than the mass of corresponding extremal BH. We plot the extremal BH mass with varying $q$ and $B$ in Fig.~\ref{extremalBHmass}.
The entropy can be calculated as
\begin{equation}
S=\int^{r_h}_0\frac{1}{T}\left(\frac{\partial M}{\partial r_h}\right)dr_h=\pi r_h^2.\label{BHentropy}
\end{equation}
Eq.~(\ref{BHentropy}) shows that the entropy does not depend on the CDF parameters directly, whereas the CDF contributes to the entropy by affecting the horizon radius $r_h$.

In the extended phase space, one can treat the cosmological constant as thermodynamic pressure and its conjugate quantity as thermodynamic volume. The definitions are as follows
\begin{eqnarray}
P&=&-\Lambda,\label{BHpressure}
\\
V&=&\left(\frac{\partial M}{\partial P}\right)_{S,q}.\label{BHthermovolume}
\end{eqnarray}
With Eq.~(\ref{BHpressure}), the mass can be reexpressed as
\begin{equation}
M=\frac{r_h}{2}+\frac{P r_h^3}{6}-\frac{r_h^3}{6}\sqrt{B+\frac{q^2}{r_h^6}}+\frac{q}{6}{\rm ArcSinh}\frac{q}{\sqrt{B}r_h^3}.\label{BHMass2}
\end{equation}
Utilizing Eqs. (\ref{BHthermovolume}) and (\ref{BHMass2}), one can obtain the thermodynamic volume as
\begin{equation}
V=\frac{r_h^3}{6}.\label{BHthermovolume2}
\end{equation}
Comparing Eqs. (\ref{BHentropy}) and (\ref{BHthermovolume2}) with those of RN-AdS BHs~\cite{Kubiznak:2012wp} and charged AdS BHs with quintessence dark energy~\cite{Li:2014ixn}, it reflects again that the expressions for both the entropy and the volume of the BH are not affected by the existence of matter content.

Since the thermodynamical behavior of a BH is highly affected by the variation of $q$, it is reasonable to treat it as a variable in the first law of thermodynamics and the Smarr relation. With the newly defined thermodynamic quantities, the first law of BH thermodynamics in the extended phase space can be written as
\begin{equation}
dM=TdS+VdP+\Phi dq,\label{ftl}
\end{equation}%
where $\Phi$ is the physical quantity conjugate to the normalization factor $q$ of CDF, and it is introduced to make the first law consistent with the Smarr relation and its physical meaning needs further investigation. Utilizing Eqs. (\ref{BHMass2}) and (\ref{ftl}), one can obtain
\begin{equation}
\Phi=\left(\frac{\partial M}{\partial q}\right)_{S,P}=\frac{1}{6}{\rm ArcSinh}\frac{q}{\sqrt{B}r_h^3}.\label{98}
\end{equation}

Considering the dimensional analysis, $[M]=1$, $[\Lambda]=-2$, $[S]=2$, $[B]=-4$ and $[q]=1$, the Smarr relation can be derived by using the Euler's theorem as
\begin{equation}
M=2TS-2VP+\Phi q.\label{100}
\end{equation}

\section{$P-V$ criticality of AdS BHs surrounded by CDF}
\label{section3}
As shown by Kubiznak and Mann~\cite{Kubiznak:2012wp}, if one treats the cosmological constant as a thermodynamic pressure, a charged AdS BH displays analogical critical behavior with the Van der Waals liquid/gas system. This kind of $P-V$ criticality maintains in the AdS BHs with the presence of Born-Infeld field~\cite{Zou:2013owa}, power Maxwell source~\cite{Hendi:2012um}, power Yang-Mills field~\cite{Biswas:2022qyl}, quintessence dark energy~\cite{Li:2014ixn}, as well as joint occurrence of Maxwell and Yang-Mills fields~\cite{Zhang:2014eap}, Born-Infeld field and quintessence~\cite{Wu:2018meo}, coupled dilaton field and Maxwell field~\cite{Dehghani:2014caa}, quintessence and cloud of strings~\cite{Chabab:2020ejk}. For the CDF-surrounded AdS BHs presented in the current paper, it is attractive for us to examine its thermodynamical phase transition by using $P-V$ criticality.
\begin{figure*}[htb]
\begin{tabular}{ c c c }
\includegraphics[width=0.43\linewidth]{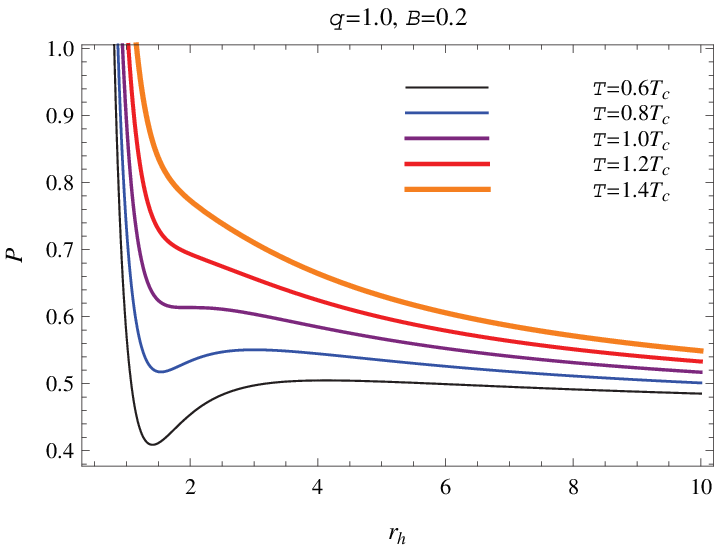}
\hspace{.10in}
\includegraphics[width=0.44\linewidth]{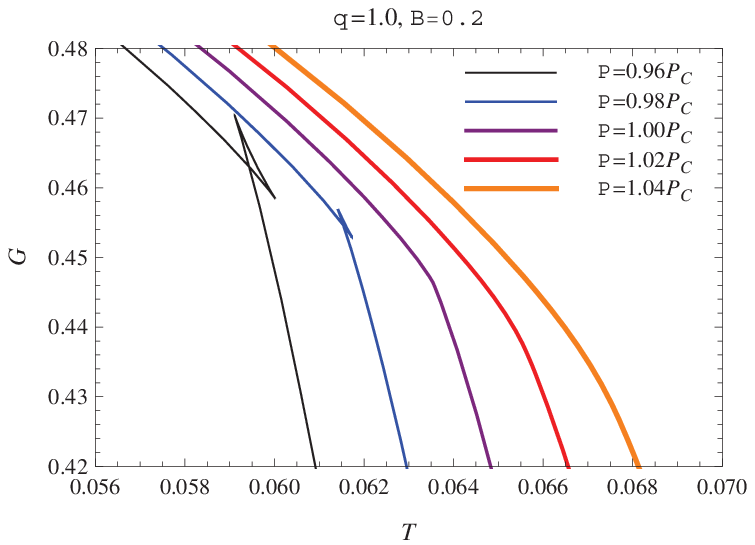}
\end{tabular}
\caption{\label{Pvcriticality} Left panel: The behavior of isothermal $P-V(r_h)$. Right panel: The behavior of isobaric $G-T$. The swallow-tail shape appears when $P<P_c$.}
\end{figure*}
\begin{table*}[htbp]
\tabcolsep 0pt
\caption{Numerical solutions for the critical physical quantities and coefficients $\mathcal{A}_i$ in $P-V$ criticality.}
\vspace*{-12pt}
\begin{center}
\def\temptablewidth{0.8\textwidth}
{\rule{\temptablewidth}{1pt}}
\renewcommand\arraystretch{1.2}
\begin{tabular*}{\temptablewidth}{@{\extracolsep{\fill}}ccccccccc}
    $B$ & $q$ & $r_c$ & $P_c$ & $T_c$ & $\frac{P_cr_c}{T_c}$ & $\mathcal{A}_1$ & $\mathcal{A}_3$ & $\mathcal{A}_5$  \\   \hline
    0.2 & 1.0 & 1.98018 & 0.61349 & 0.06353 & 19.1236 & 0.65711 & -0.21904 & -0.01798  \\
    0.3 & 1.0 & 1.88657 & 0.73135 & 0.06676 & 20.6663 & 0.60806 & -0.20269 & -0.01685  \\
    0.4 & 1.0 & 1.82257 & 0.82950 & 0.06916 & 21.8586 & 0.57490 & -0.19163 & -0.01606  \\
    0.2 & 0.8 & 1.76045 & 0.65626 & 0.07121 & 16.2247 & 0.77452 & -0.25817 & -0.02048  \\
    0.2 & 1.2 & 2.17761 & 0.58527 & 0.05789 & 22.0144 & 0.57082 & -0.19028 & -0.01596  \\
\end{tabular*}
{\rule{\temptablewidth}{1pt}}
\end{center}
\label{pvplane}
\end{table*}
\begin{figure*}[htb]
\begin{tabular}{ c c c }
\includegraphics[width=0.43\linewidth]{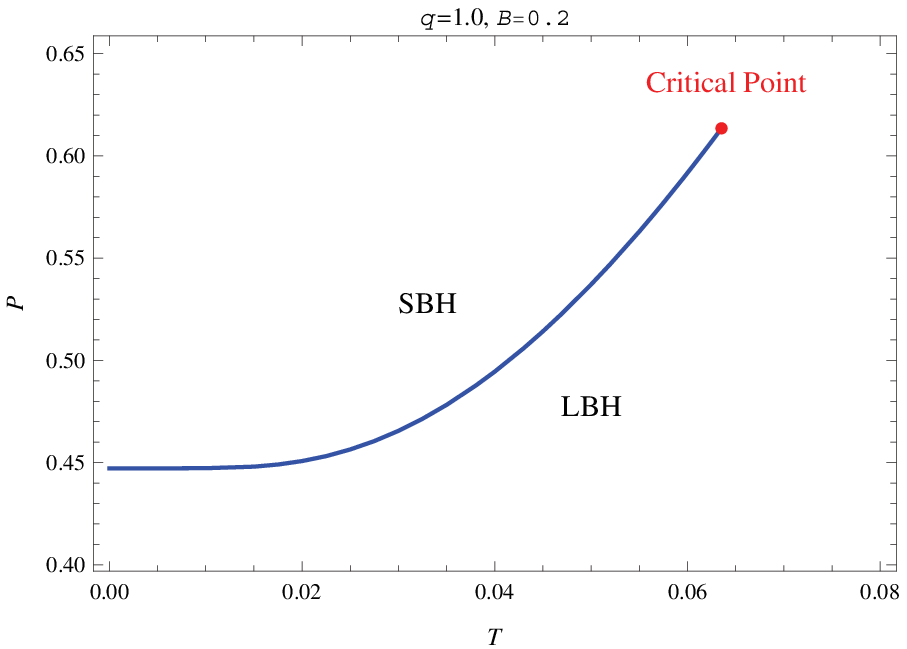}
\hspace{.10in}
\includegraphics[width=0.43\linewidth]{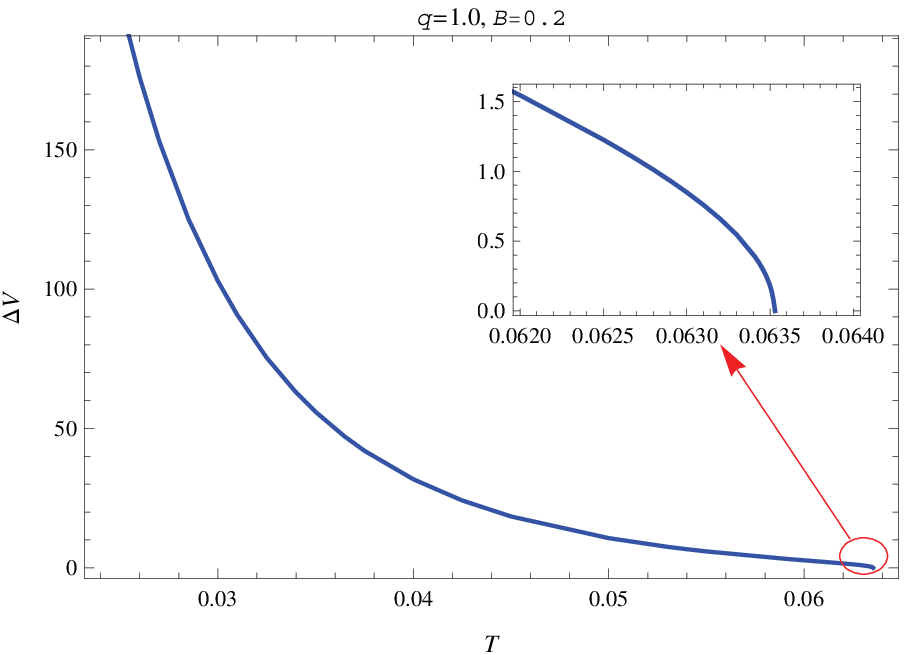}
\end{tabular}
\caption{\label{Pvphasestructrue} Left panel: Coexistence curve of small/large BH phase transition in the $P-T$ plane. Right panel: The $\Delta V$ as function of temperature $T$. The fitting $\Delta V-T$ curve near the critical temperature is magnified into view.}
\end{figure*}
\subsection{$P-V$ criticality}
With Eqs. (\ref{BHtemperature}) and (\ref{BHpressure}), the Hawking temperature can be reexpressed as
\begin{equation}
T=\frac{1}{4\pi}\left(\frac{1}{r_h}+P r_h-r_h\sqrt{B+\frac{q^2}{r_h^6}}\right).\label{12}
\end{equation}
From Eq. (\ref{12}), one can easily derive the equation of state of the BH as
\begin{equation}
P=\frac{4\pi}{r_h}T-\frac{1}{r_h^2}+\sqrt{B+\frac{q^2}{r_h^6}},\label{13}
\end{equation}
where the event horizon $r_h$ is related to the thermodynamic volume $V$ by Eq.~(\ref{BHthermovolume2}). The third term in the above expression for $P$ reflects the effect of the CDF. Note that the pressure $P$ in extended phase space thermodynamics of asymptotically AdS BHs is entirely due to the variable cosmological constant, according to Eq.~(\ref{BHpressure}), thus the pressure of the matter fluid should appear in the equation of state of the BH. This is different from the case in horizon thermodynamics context~\cite{Hansen:2016ayo}, where $P$ is identified with the total pressure of all matter in the spacetime, including the cosmological constant.

The critical point can be derived through the following conditions
\begin{eqnarray}
\left.\frac{\partial P}{\partial r_h}\right|_{T=T_c}&=&0,\label{15}\\
\left.\frac{\partial ^2P}{\partial r_h^2}\right|_{T=T_c}&=&0.\label{16}
\end{eqnarray}
Utilizing  Eqs. (\ref{13}) and (\ref{15}), one can obtain
\begin{equation}
T_c=\frac{1}{2\pi r_c}-\frac{3q^2}{4\pi\rho\left(r_c\right)r_c^5},\label{17}
\end{equation}
where $T_c$ and $r_c$ denote the critical Hawking temperature and critical event horizon radius respectively, $\rho\left(r_c\right)$ denotes the function value of $\rho(r)$ at $r=r_c$.
Utilizing Eqs. (\ref{13}), (\ref{16}) and (\ref{17}), the condition that $r_c$ satisfies can be derived as
\begin{equation}
6q^4+15Bq^2r_c^6-2\rho\left(r_c\right)^3r_c^{10}=0.\label{18}
\end{equation}
By using Eqs. (\ref{13}) and (\ref{17}), one can obtain the critical pressure as
\begin{equation}
P_c=\frac{1}{r_c^2}-\frac{3q^2}{\rho\left(r_c\right)r_c^6}+\rho\left(r_c\right).\label{102}
\end{equation}

To study the dependencies of the critical physical quantities on the CDF parameters, we appeal to numerical method for help. To observe the influence of the parameters respectively, one can let one parameter vary while keeping the other one fixed. For specific values of parameters, Eq.~(\ref{18}) can be solved numerically by Mathematica programming and the critical horizon radius can be derived. Then the critical temperature can be obtained through Eq.~(\ref{17}) and the critical pressure can be derived through Eq.~(\ref{102}).

The corresponding critical physical quantities are shown in Table~\ref{pvplane}. Fixing $B=0.2$ and varying $q$ from $0.8$ to $1.2$, one finds that the critical volume (positively related to $r_c$) increases with $q$ while both the critical Hawking temperature $T_c$ and critical pressure $P_c$ decrease with it. We set $q=1$ and let $B$ vary from $0.2$ to $0.4$, finding that $r_c$ decreases with $B$ while both $T_c$ and $P_c$ increase with it. For the ratio $\frac{P_cr_c}{T_c}$, it does not keep constant as the charged AdS BH does~\cite{Kubiznak:2012wp}, one can observe from Table~\ref{pvplane} that the ratio increases with $q$, as well as $B$. This phenomenon reflects the effects of the CDF.

To observe the behavior of $P$ more intuitively, its curve is plotted under different temperature. As shown in the left panel of Fig.~\ref{Pvcriticality}, the two upper lines for $T>T_c$ correspond to the ``ideal gas" one-phase behavior, thus no phase transition occurs. The critical isotherm at $T=T_c$ has an inflection point, indicating the occurrence of a second-order phase transition. The isotherm corresponding to the temperature less than the critical temperature can be divided into three branches. Both the small radius branch and the large radius branch are stable while the medium radius branch is unstable. There is a first-order phase transition between the small BH and the large BH. As discussed in Ref.~\cite{Kubiznak:2012wp}, to describe this phase transition, one has to replace the `oscillating' part of the isotherm by an isobar. It has been demonstrated that the Maxwell's equal area law is valid for $P-V$ diagram~\cite{Wei:2014qwa}, yielding
\begin{equation}\label{Maxwellequalarea}
\oint V dP=0,
\end{equation}
or equivalently reexpressed as
\begin{equation}\label{Maxwellequalareav2}
\int_{V_s}^{V_l}PdV=P^*\left(V_l-V_s\right),
\end{equation}
where $P^*$ denotes the pressure at which the phase transition occurs, $V_l$ and $V_s$ denote the thermodynamic volumes of large BH and small BH respectively. The Maxwell's equal area is an effective tool to find this coexistence pressure, or to calculate the change of the volume between the two phases of small and large BHs, i.e. $\Delta V=V_l-V_s$. With the help of Eq.~(\ref{Maxwellequalareav2}), we numerically plot the coexistence curve and the $\Delta V-T$ curve in Fig.~\ref{Pvphasestructrue}. The critical point is highlighted by a small circle at the end of the coexistence line. $\Delta V$ is a monotone decreasing function of $T$, it decreases to zero at the critical temperature $T_c$.

To understand this phase transition more deeply, one can analyze the behavior of Gibbs free energy. In the extended phase space, the mass is interpreted as enthalpy. So the Gibbs free energy can be derived as
\begin{eqnarray}
G&=&H-TS=M-TS~\nonumber \\
&=&\frac{r_h}{4}-\frac{P r_h^3}{12}+\frac{r_h^3}{12}\rho(r_h)+\frac{q}{6}{\rm ArcSinh}\frac{q}{\sqrt{B}r_h^3}.\label{20}
\end{eqnarray}
The behavior of Gibbs free energy is depicted in the second panel of Fig.~\ref{Pvcriticality}. The classical swallow tail phenomenon observed in the $G-T$ curve below the critical pressure implies the existence of first-order phase transition.

\subsection{Critical exponents for $P-V$ criticality}
Critical exponents are often used to describe the critical behavior near the critical point. It is convenient to introduce the following notations
\begin{equation}
t=\frac{T}{T_c}-1,\;\;\epsilon=\frac{V}{V_c}-1,\;\;p=\frac{P}{P_c},\label{21}
\end{equation}
where the critical thermodynamic volume $V_c$ is related to the critical event horizon radius $r_c$ by $V_c=\frac{r_c^3}{6}$ . The definitions of critical exponents are as follows
\begin{eqnarray}
C_V&\propto&|t|^{-\alpha},\label{22}\\
\eta&\propto&|t|^{\beta},\label{23}\\
\kappa_{T@\langle P-V\rangle}&\propto&|t|^{-\gamma},\label{24}\\
|P-P_c|&\propto&|V-V_c|^{\delta}.\label{25}
\end{eqnarray}

$\alpha$ describes the behavior of specific heat when the volume is fixed. From Eq. (\ref{BHentropy}), it is not difficult to draw the conclusion that the entropy $S$ is independent of the Hawking temperature $T$, so
\begin{equation}
C_V=T\left(\frac{\partial S}{\partial T}\right)_V=0,\label{26}
\end{equation}
from which one can derive that $\alpha=0$.

$\beta$ characterizes the behavior of the order parameter $\eta$. Near the critical point, the equation of state can be expanded into
\begin{equation}
p=1+\mathcal{A}_1t+\mathcal{A}_2\epsilon+\mathcal{A}_3t\epsilon+\mathcal{A}_4\epsilon^2+\mathcal{A}_5\epsilon^3+O(t\epsilon^2,\epsilon^4),\label{27}
\end{equation}
where
\begin{eqnarray}
\mathcal{A}_1&=&\frac{4\pi T_c}{P_cr_c},\label{29}\\
\mathcal{A}_2&=&\mathcal{A}_4=0,\label{28}\\
\mathcal{A}_3&=&-\frac{4\pi T_c}{3P_cr_c},\label{30}\\
\mathcal{A}_5&=&-\frac{56\pi T_c}{81P_cr_c}+\frac{40}{81 P_cr_c^2}-\frac{q^2A}{2 \rho\left(r_c\right)^5 P_c r_c^{18}},\label{31}
\end{eqnarray}
with $A=2q^4+5Bq^2r_c^6+4B^2r_c^{12}$. The dependencies of coefficients $\mathcal{A}_i$ on parameters $B$ and $q$ can be found by the numerical results presented in Table~\ref{pvplane}.

Since the pressure keeps constant during the phase transition, one can obtain
 \begin{equation}
1+\mathcal{A}_1t+\mathcal{A}_3t\epsilon_l+\mathcal{A}_5\epsilon_l^3=1+\mathcal{A}_1t+\mathcal{A}_3t\epsilon_s+\mathcal{A}_5\epsilon_s^3.\label{32}
\end{equation}
On the other hand, one can apply Maxwell's equal area law
\begin{equation}
\int^{\epsilon_s}_{\epsilon_l}\epsilon \frac{dp}{d\epsilon}d \epsilon=0,\label{33}
\end{equation}
with
\begin{equation}
\frac{dp}{d\epsilon}=\mathcal{A}_3t+3\mathcal{A}_5\epsilon^2,\label{34}
\end{equation}
and obtain
\begin{equation}
\mathcal{A}_3t(\epsilon^2_s-\epsilon^2_l)+\frac{3}{2}\mathcal{A}_5(\epsilon^4_s-\epsilon^4_l)=0.\label{35}
\end{equation}
From Eqs. (\ref{32}) and (\ref{35}), one can obtain
\begin{equation}
\epsilon_l=-\epsilon_s=\sqrt{-\frac{\mathcal{A}_3}{\mathcal{A}_5}t},\label{36}
\end{equation}
where the argument under the square root function keeps positive, considering that the small/large BH phase transition occurs when $T<T_c$ and $\mathcal{A}_{3}$ has the same sign with $\mathcal{A}_{5}$ according to the numerical results presented in Table~\ref{pvplane}. So
\begin{equation}
\eta=V_l-V_s=V_c(\epsilon_l-\epsilon_s)=2V_c\epsilon_l\propto\sqrt{-t},\label{37}
\end{equation}
yielding $\beta=1/2$. This behavior of $V_l-V_s$ near the critical point can also be revealed by the magnified view of $\Delta V-T$ curve near $T_c$ in the right panel of Fig.~\ref{Pvphasestructrue}.

$\gamma$ describes the behavior of isothermal compressibility $\kappa_{T@\langle P-V\rangle}$, which can be derived as
\begin{equation}
\kappa_{T@\langle P-V\rangle}=\left.-\frac{1}{V}\frac{\partial V}{\partial P}\right|_{V_c}=\left.-\frac{1}{P_c}\frac{1}{\frac{\partial p}{\partial \epsilon}}\right|_{\epsilon=0}\propto \frac{r_c}{T_c}t^{-1},\label{38}
\end{equation}
which yields $\gamma=1$.

$\delta$ characterizes the behavior described in Eq.~(\ref{25}) on the critical isotherm $T=T_c$. Substituting $t=0$ into Eq. (\ref{27}), one can obtain
\begin{equation}
|P-P_c|=P_c|p-1|=P_c\left|\mathcal{A}_5\epsilon^3\right|=\frac{P_c\left|\mathcal{A}_5\right|}{V_c^3}|V-V_c|^3,\label{39}
\end{equation}
yielding $\delta=3$.

From the above derivations, we can see clearly that four critical exponents are exactly the same as those obtained before for charged AdS BHs~\cite{Kubiznak:2012wp}. This implies that the CDF does not change the critical exponents, just like the quintessence dark energy~\cite{Li:2014ixn}. The universality of van der Waals like phase transition, as well as the values of critical exponents, for AdS BHs, has been further verified.

\section{Critical behavior of AdS BHs surrounded by CDF in $q-\Phi$ phase space with a fixed cosmological constant}
\label{section4}
Niu, Tian and Wu~\cite{Niu:2011tb} studied the phase transitions and critical phenomena for the Reissner-Nordstrom (RN) BH in ($n+1$)-dimensional AdS spacetime, and found that near the critical point, the $Q-\Phi$ diagram shares the same shape as that of $P-V$ diagram for a van der Waals liquid/gas system, which strongly suggests a remarkable analogy between these two thermodynamic systems. Zhou and Wei~\cite{Zhou:2019xai} studied the charge-electric potential criticality for the charged AdS BHs with carefully investigating the equal area law. Very recently, Hendi and Jafarzade~\cite{Hendi:2020ebh} found that, charged AdS BHs surrounded by quintessence admits the same criticality and van der Waals like behavior, by considering the normalization factor which indicates the intensity of the quintessence field, as a thermodynamic variable. For an AdS BH surrounded by CDF, we have shown there exists a small/large BH phase transition in $P-V$ phase space, analogous to the liquid/gas phase transition of Van der Waals fluids. It will be interesting to probe the phase transition and critical behavior by treating the normalization factor $q$ as a thermodynamic variable and keeping the cosmological constant as fixed parameter.
\begin{figure*}[htbp]
\begin{tabular}{ c c c }
\includegraphics[width=0.42\linewidth]{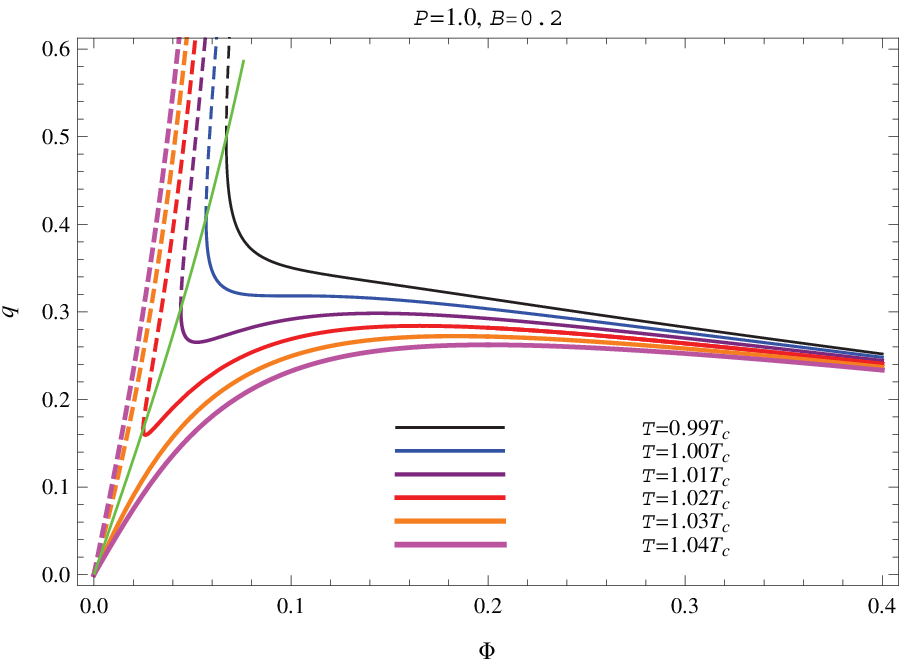}
\hspace{.10in}
\includegraphics[width=0.45\linewidth]{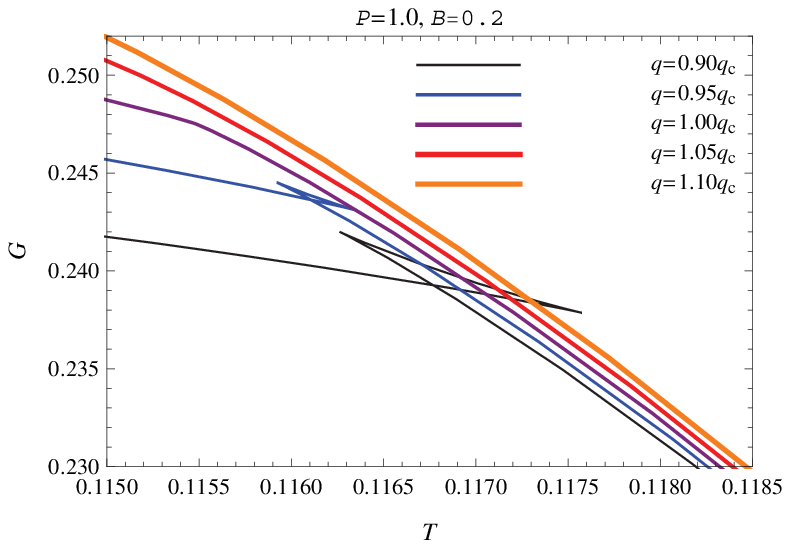}
\end{tabular}
\caption{\label{qphicriticality} Left panel: The behavior of isothermal $q-\Phi$. The dashed and solid lines are for $q_1$ and $q_2$, respectively. The thin green line denotes the connection point $\left(\Phi_0,q_0\right)$ of the curves $q_1$ and $q_2$. Right panel: Gibbs free energy $G$ as a function of temperature $T$ for fixed $q$. The swallow-tail shape appears when $q<q_c$. The critical quantities corresponding to $B=0.2$ and $P=1.0$ are $\Phi_c=0.09751$, $q_c=0.31829$ and $T_c=0.11548$.}
\end{figure*}
\subsection{$q-\Phi$ criticality with a fixed cosmological constant}
In this subsection, we consider the normalization factor $q$ as a thermodynamic variable and study the critical behavior of the system under its variation. We start by rewriting the relation of temperature. Inserting Eq. (\ref{98}) into Eq.~(\ref{12}), one obtains
\begin{eqnarray}
T=\frac{1}{4\pi}\left(\frac{P q^{1/3}}{\Xi(\Phi)}+\frac{\Xi(\Phi)}{q^{1/3}}-\frac{q^{1/3}\sqrt{B+\Xi(\Phi)^6}}{\Xi(\Phi)}\right),  \label{EqaTy}
\end{eqnarray}
with $\Xi(\Phi)\equiv B^{1/6}({\rm Sinh}6\Phi)^{1/3}$. Solving $q$ from the above equation, we obtain two solutions for the BH equation of state
\begin{eqnarray}
q_1 &=&\left(\frac{-\sqrt{64\pi^2T^2+8\Upsilon(\Phi)}-8\pi T}{2\Upsilon(\Phi)}\Xi(\Phi)\right)^3,  \\
&&  \nonumber \\
q_2 &=&\left(\frac{\sqrt{64\pi^2T^2+8\Upsilon(\Phi)}-8\pi T}{2\Upsilon(\Phi)}\Xi(\Phi)\right)^3,  \label{EqaTc}
\end{eqnarray}
with $\Upsilon(\Phi)\equiv-2P+2\sqrt{B}{\rm Cosh}6\Phi$. So $q$ is a double-valued function of $\Phi$, while $q_1$ and $q_2$ are single-valued functions. Moreover, $q_1$ and $q_2$ meet each other at
\begin{eqnarray}
\Phi_0 &=&\frac{1}{12}{\rm ArcCosh}\left[\frac{2\left(-P+4\pi^2 T^2\right)^2}{B}-1\right],  \nonumber \\
&&  \nonumber \\
q_0 &=&\left(\frac{\Xi(\Phi_0)}{2\pi T}\right)^3,  \label{phi01}
\end{eqnarray}
when $T<\frac{1}{2\pi}\sqrt{P-\sqrt{B}}$, while they meet at
\begin{equation}
\Phi_0=0,~~~~q_0=0,\label{phi02}
\end{equation}
when $T>\frac{1}{2\pi}\sqrt{P-\sqrt{B}}$.

In Fig.~\ref{qphicriticality}, we plot $q$ as a function of $\Phi$ with fixing $P$. It is obvious that $q_{1}$ always increases with $\Phi$. While for $q_{2}$, it has a different behavior. If the temperature $T<T_{\rm c}$, $q_{2}$ decreases with $\Phi$. However, when $T_{\rm c}<T<\frac{1}{2\pi}\sqrt{P-\sqrt{B}}$, $q_{2}$ first decreases, then increases, and finally decreases with $\Phi$. While when $T>\frac{1}{2\pi}\sqrt{P-\sqrt{B}}$,  $q_{2}$ firstly increases, then decreases with $\Phi$. As a result, an isotherm for $T>T_c$ in $q-\Phi$ plane has a local maxima and a minima, between which there is a segment similar to the `oscillating' part of the isotherm for $T<T_c$ in $P-V$ plane, thus indicating the existence of a first-order phase transition between low-$\Phi$ and high-$\Phi$ BHs. The $G-T$ curves for varying normalization factor in $q-\Phi$ criticality are also depicted in Fig.~\ref{qphicriticality}, and the emergent swallow-tail for $q<q_c$ also denotes the existence of Van der Waals like phase transition.

When we consider the critical point in the $q-\Phi$ plane, it is natural to use Eq.~(\ref{EqaTc}) and the concept of the inflection point to characterize the critical point by
\begin{equation}
\frac{\partial q_2}{\partial \Phi }\bigg|_{\Phi =\Phi_{c},T=T_{c}}=0~~~\&~~~%
\frac{\partial ^{2}q_2}{\partial \Phi ^{2}}\bigg|_{\Phi =\Phi_{c},T=T_{c}}=0.
\label{EqCriticalpoint}
\end{equation}
However, one finds that it is difficult to tackle with these equations and obtain analytical solutions for the critical quantities. Keeping in mind that, the phase structure of a thermodynamic system can be characterized by the Gibbs free energy, both $P$ and $q$ can change the $G-T$ behavior, thus the phase structure, at the critical point, there should exist a one-to-one correlation between the critical quantities in the $P-V$ criticality and $q-\Phi$ criticality. That is to say
\begin{equation}\label{correspond}
\mathop{\overbrace{B, q, P_c, r_c, T_c}}^{P-V~{\rm criticality}}\Longleftrightarrow\mathop{\overbrace{B, P, q_c, \Phi_c, T_c}}^{q-\Phi~{\rm criticality}}.
\end{equation}
The critical normalization factor and temperature corresponding to $B=0.2$ and $P=1.0$ are numerically calculated as $q_c=0.31829$ and $T_c=0.11548$, the $q-\Phi$ plot and $G-T$ plot in Fig.~\ref{qphicriticality} verify the correlation discussed above. This conclusion can also be confirmed by observing the heat capacity of the system in both phase spaces. The heat capacity in $P-V$ criticality with fixed $q$ is calculated as
\begin{figure*}[htb]
\begin{tabular}{ c c }
\includegraphics[width=0.43\linewidth]{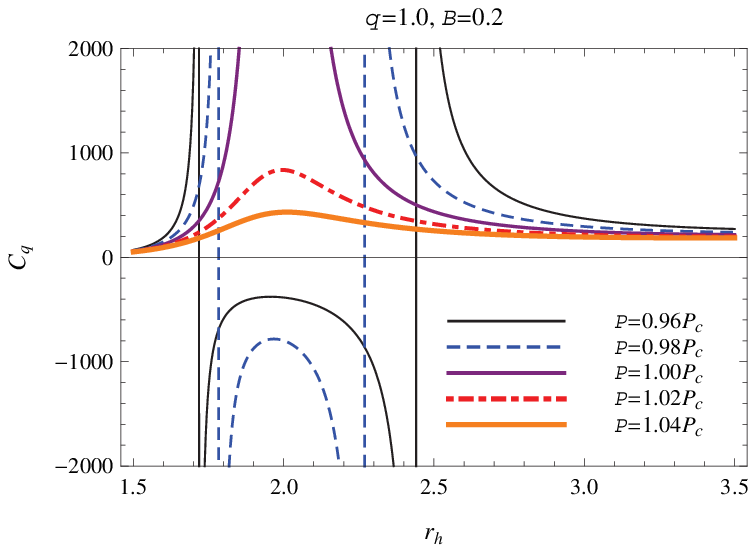}
\hspace{.10in}
\includegraphics[width=0.43\linewidth]{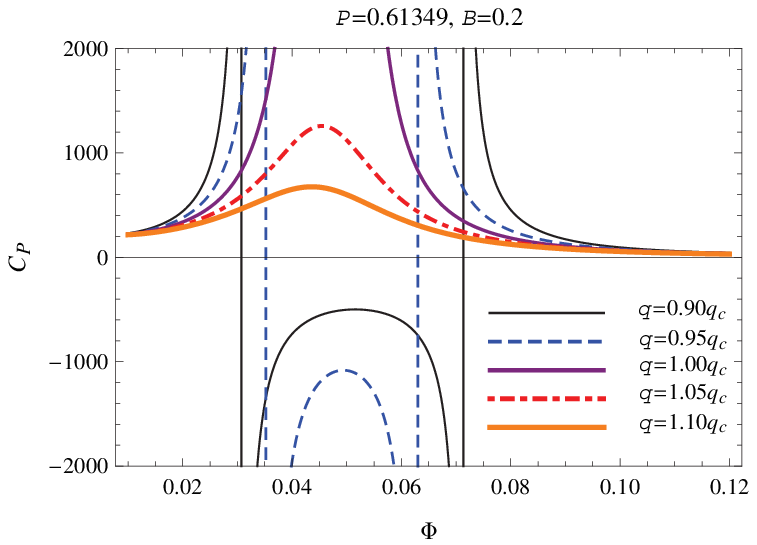}
\end{tabular}
\caption{\label{Cvphi} The behaviors of the heat capacity in $P-V$ criticality (left panel) and $q-\Phi$ criticality (right panel). The critical normalization factor in the $q-\Phi$ criticality, $q_c=1.0$ when $B=0.2~\&~P=0.61349$, corresponds to the critical pressure in the $P-V$ criticality, $P_c=0.61349$ when $B=0.2~\&~q=1.0$.}
\end{figure*}
\begin{eqnarray}
C_q&=&T\left(\frac{\partial S}{\partial T}\right)_q=T\left(\frac{\partial S}{\partial V}\frac{\partial V}{\partial T}\right)_q~\nonumber \\
&=&\frac{\pi r_h^2 \mathcal{I}(r_h)\left(2r_h+2Pr_h^3-2\mathcal{I}(r_h)\right)}{2q^2-B r_h^6+r_h\left(-1+Pr_h^2\right)\mathcal{I}(r_h)},\label{Cq}
\end{eqnarray}
with $\mathcal{I}(r_h)=\sqrt{q^2+Br_h^6}$, and the heat capacity in $q-\Phi$ criticality with fixed cosmological constant is expressed as
\begin{eqnarray}
C_P&=&T\left(\frac{\partial S}{\partial T}\right)_P=T\left(\frac{\partial S}{\partial \Phi}\frac{\partial \Phi}{\partial T}\right)_P~\nonumber \\
&=&\frac{2\pi \mathcal{J}(\Phi)\mathrm{Cosh}6\Phi \left(\mathrm{Csch}6\Phi\right)^{2/3}}{B^{5/6}\left(-2+\mathrm{Cosh}12\Phi\right)-\mathcal{K}(\Phi)\mathrm{Cosh}6\Phi },\label{CP}
\end{eqnarray}
with
\begin{eqnarray}
\mathcal{J}(\Phi)&=& P q^{2/3}-\sqrt{B}q^{2/3}\mathrm{Cosh}6\Phi+\Xi(\Phi)^2,~\nonumber \\
\mathcal{K}(\Phi)&=&-PB^{1/3}+\left(q^{-1}B\mathrm{Sinh}6\Phi\right)^{2/3}.\label{JK}
\end{eqnarray}
The behaviors of the heat capacity in $P-V$ criticality and $q-\Phi$ criticality are depicted in Fig.~\ref{Cvphi}. As well known, the heat capacity provides the information related to the thermal stability and phase transition of a thermodynamic system. The sign of heat capacity determines thermal stability/instability of BHs. The positivity (negativity) of this quantity indicates a BH is thermally stable (unstable). What's more, the phase transition points are where heat capacity diverges, and the divergence, or to say discontinuity, disappears exactly when the pressure (in $P-V$ criticality) or normalization factor (in $q-\Phi$ criticality) reach their critical values. One can observe from Fig.~\ref{Cvphi} that, the critical quantities in one phase space can lead to critical quantities in the other one. Based on the one-to-one correlation between the critical quantities in $P-V$ and $q-\Phi$ phase spaces, as discussed above, the effects of the parameters $B$ and $P$ on critical quantities in $q-\Phi$ phase space can be studied numerically. The numerical results in Table~\ref{qphiplane} show that, $q_c$ increases with $B$ while decreases with $P$, both $\Phi_c$ and $T_c$ decrease with $B$ while increase with $P$.\begin{table*}[htbp]
\tabcolsep 0pt
\caption{Numerical solutions for the critical physical quantities and coefficients $\mathcal{B}_i$ in $q-\Phi$ criticality.}
\vspace*{-12pt}
\begin{center}
\def\temptablewidth{0.9\textwidth}
{\rule{\temptablewidth}{1pt}}
\renewcommand\arraystretch{1.2}
\begin{tabular*}{\temptablewidth}{@{\extracolsep{\fill}}cccccccccc}
        $B$ & $P$ & $\Phi_c$ & $q_c$ & $T_c$ & $\mathcal{B}_1$ & $|\mathcal{B}_2|$ & $\mathcal{B}_3$ & $|\mathcal{B}_4|$ & $\mathcal{B}_5$ \\   \hline
        0.2 & $1.0$ & 0.09751 & 0.31829 & 0.11548 & -9.50686 & $<10^{-10}$ & 14.68550 & $<10^{-10}$ & -0.24436 \\
        0.3 & $1.0$ & 0.07481 & 0.41852 & 0.10462 & -10.40520 & $<10^{-10}$ & 16.52530 & $<10^{-10}$ & -0.28901 \\
        0.4 & $1.0$ & 0.06074 & 0.54486 & 0.09439 & -10.90300 & $<10^{-10}$ & 17.58190 & $<10^{-10}$ & -0.31421 \\
        0.2 & $0.9$ & 0.08513 & 0.38250 & 0.10461 & -10.00730 & $<10^{-10}$ & 15.69950 & $<10^{-10}$ & -0.26907 \\
        0.2 & $1.1$ & 0.11035 & 0.27410 & 0.12536 & -8.97503 & $<10^{-10}$ & 13.63780 & $<10^{-10}$ & -0.21860 \\
       \end{tabular*}
       {\rule{\temptablewidth}{1pt}}
       \end{center}
       \label{qphiplane}
       \end{table*}
\begin{figure*}[htb]
\begin{tabular}{ c c c }
\includegraphics[width=0.43\linewidth]{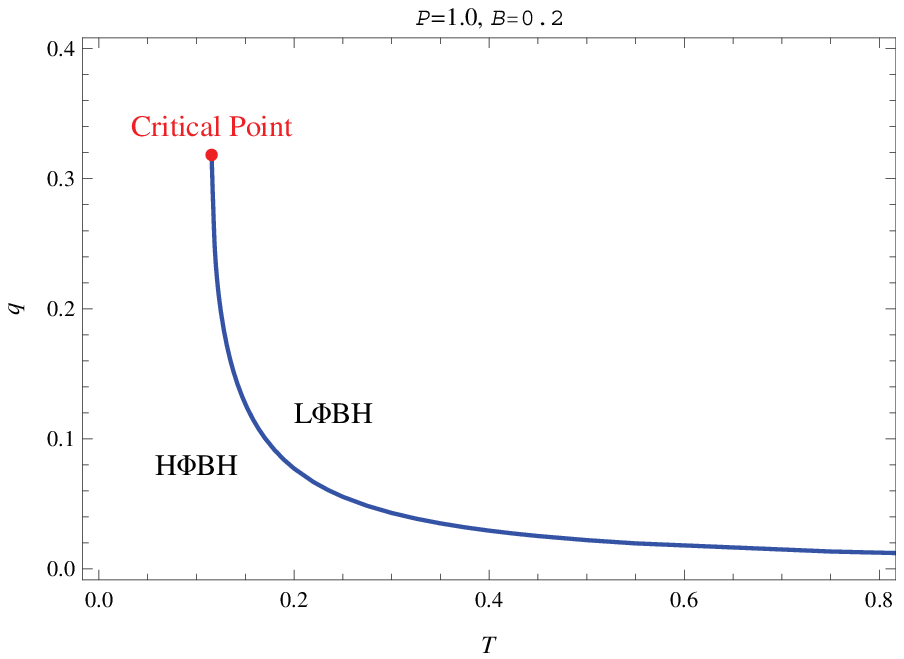}
\hspace{.10in}
\includegraphics[width=0.432\linewidth]{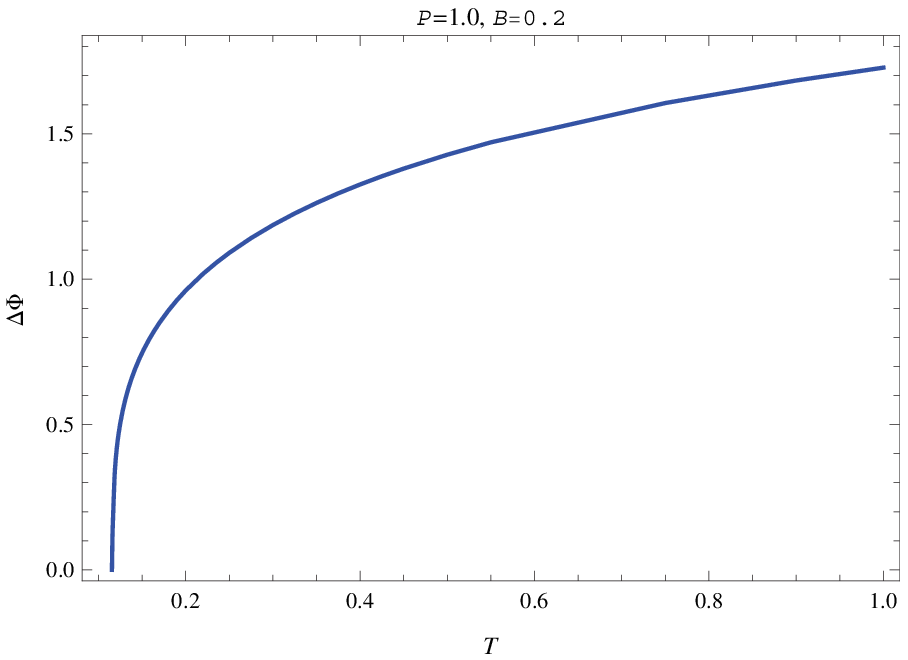}
\end{tabular}
\caption{\label{qphiphasestructure} Left panel: Coexistence curve of low/high-$\Phi$ BH phase transition in the $q-T$ plane. Right panel: The $\Delta \Phi$ as function of temperature $T$.}
\end{figure*}
To describe the low/high-$\Phi$ phase transition which is of first-order and find its terminating
point, let us first rewrite the first law expressed in Eq.~(\ref{ftl}) via a Legendre transformation
\begin{equation}\label{Legendrefirst}
dG=-SdT+VdP+\Phi dq.
\end{equation}
Note that a first-order transition in the $q-\Phi$ phase space occurs at a fixed temperature and normalization factor, and both the coexistence phases have the same Gibbs free energy. Thus, one has
\begin{equation}\label{Legendrefirstv2}
0=VdP+\Phi dq.
\end{equation}
If one works in a canonical ensemble with fixed $P$, then the condition in Eq.~(\ref{Legendrefirstv2}) leads to the equal area law in $q-\Phi$ plane
\begin{equation}\label{EALq_phi}
\oint \Phi dq=0.
\end{equation}
To change the integral variable from $q$ to $\Phi$ in Eq.~(\ref{EALq_phi}), one should keep in mind that $q$ is a double-valued function of $\Phi$. By using the techniques developed by Zhou and Wei in Ref.~\cite{Zhou:2019xai}, the equal area law in Eq.~(\ref{EALq_phi}) can be reexpressed as
\begin{equation}\label{EALq_phi1}
\int_{\Phi_L}^{\Phi_H} q_2d\Phi=q^*\left(\Phi_H-\Phi_L\right),
\end{equation}
near the critical point, and
\begin{equation}\label{EALq_phi2}
\int_{\Phi_L}^{\Phi_0} q_1d\Phi+\int_{\Phi_0}^{\Phi_H} q_2d\Phi=q^*\left(\Phi_H-\Phi_L\right),
\end{equation}
far from the critical point, where $q^*$ denotes the coexistence normalization factor at which the first-order phase transition occurs, the subscripts $H$ and $L$ represents the high-$\Phi$ and low-$\Phi$ BHs, respectively. By using the equal area law expressed in Eqs.~(\ref{EALq_phi1}) and~(\ref{EALq_phi2}), we numerically plot the coexistence curve of low/high-$\Phi$ BH phase transition, as well as the $\Delta\Phi-T$ curve in Fig.~\ref{qphiphasestructure}. Although the equal area laws have two different expressions near the critical point and far from the critical point, their coexistence curves are smoothly connected. $\Delta\Phi$ is a monotone increasing function of $T$, this is due to the fact that the first-order low/high-$\Phi$ BH phase transition occurs when $T>T_c$.

\subsection{Critical exponents for $q-\Phi$ criticality}
Now, we aim to calculate the critical exponents for $q-\Phi$ criticality. We first introduce the following relations
\begin{eqnarray}
C_{\Phi}&\propto&|t|^{-\lambda},\label{22p}\\
\vartheta&\propto&|t|^{\chi},\label{23p}\\
\kappa_{T@\langle q-\Phi\rangle}&\propto&|t|^{-\sigma},\label{24p}\\
|q_2-q_{c}|&\propto&|\Phi-\Phi_c|^{\iota}.\label{25p}
\end{eqnarray}
where the critical exponents $\lambda$ , $\chi$, $\sigma$ and $\iota$
describe the behaviors of specific heat $C_{\Phi} $, the order parameter $\vartheta$, the isothermal compressibility $\kappa_{T@\langle q-\Phi\rangle}$ and behavior on the
critical isotherm $T=T_{c}$, respectively. To find the critical exponent,
we define the below dimensionless quantities
\begin{equation}
\xi=\frac{q_2}{q_{c}},~~~\zeta=\frac{\Phi}{\Phi_{c}}-1,~~~t=\frac{T}{T_{c}}-1.
\label{Eqreduced1}
\end{equation}

With the above definition, the physical quantities can be expressed as
\begin{equation}
q_2=\xi q_c,~~\Phi=(1+\zeta)\Phi_c,~~T=(1+t)T_c.  \label{EqNereduced1}
\end{equation}

We rewrite the entropy in terms of $T $ and $\Phi$ as
\begin{equation}
S(T,\Phi)=\frac{\pi q^{2/3}}{\Xi(\Phi)^2},  \label{EqNS}
\end{equation}
which is independent of temperature. So, we find that
\begin{equation}
C_{\Phi}=T\frac{\partial S}{\partial T}\bigg|_{\Phi}=0,
\end{equation}
and hence $\lambda =0 $. By using Eq. (\ref{EqNereduced1}),
one can expand Eq. (\ref{EqaTy}) near the critical point as
\begin{equation}
\xi =1+\mathcal{B}_{1}t+\mathcal{B}_{2}\zeta+\mathcal{B}_{3}\zeta t
+\mathcal{B}_{4}\zeta^{2}+\mathcal{B}_{5}\zeta^{3}+O(t\zeta^{2},
\zeta^{4}), \label{Eqnu}
\end{equation}
where
\begin{eqnarray}
\mathcal{B}_{1}&=&-\frac{6\pi T_c}{B_1},  \nonumber \\
&&  \nonumber \\
\mathcal{B}_{2}&=&\frac{3B^{1/6}B_4\Phi_c\left({\rm Csch6\Phi_c}\right)^{2/3}}{2B_1B_2q_c^{1/3}},  \nonumber \\
&&  \nonumber \\
\mathcal{B}_{3}&=&\frac{6\pi B_7\Phi_c T_c}{B_1^3B_2^2q_c},  \nonumber \\
&&  \nonumber \\
\mathcal{B}_{4}&=&\frac{3B^{1/6}B_{11}\Phi_c^2}{16B_2^2q_c^{2/3}},  \nonumber \\
&&  \nonumber \\
\mathcal{B}_{5}&=&\frac{B_{18}\Phi_c^3}{64B_1^5B_2^3q_c^{2/3}},~~~~~\label{EqaTc1}
\end{eqnarray}
within which, the expressions for $B_i$ are at some level redundant, thus we list them in Appendix~\ref{appendix1}.

Our numerical analysis in Table \ref{qphiplane} shows that the coefficients $\mathcal{B}_{2}$ and $\mathcal{B}_{4}$ are very small and can be considered as zero. So, Eq. (\ref{Eqnu}) reduces to
\begin{equation}
\xi =1+\mathcal{B}_{1}t+\mathcal{B}_{3}\zeta t+\mathcal{B}_{5}\zeta^{3}.  \label{Eqnu2}
\end{equation}

Differentiating Eq. (\ref{Eqnu2}) with respect to $\zeta$
for a fixed $ t $, we get
\begin{equation}
dq_2= q_{c}(\mathcal{B}_{3} t+3\mathcal{B}_{5}\zeta^{2})d\zeta.  \label{Eqda}
\end{equation}

Now, using the fact that the normalization factor remains constant during the phase transition and employing the Maxwell's area law~(\ref{EALq_phi}), we have the following two equations:
\begin{eqnarray}
\xi-1 &=&  \mathcal{B}_{1}t+\mathcal{B}_{3}\zeta_{H} t+\mathcal{B}_{5}\zeta_{H}^{3}=\mathcal{B}_{1}t+\mathcal{B}_{3}\zeta_{L} t+\mathcal{B}_{5}\zeta_{L}^{3},  \nonumber \\
&&  \nonumber \\
0&=&\int_{\zeta_{L}}^{\zeta_{H}}\zeta (\mathcal{B}_{3} t+3\mathcal{B}_{5}\zeta^{2})d\zeta.
\label{EqMaxwell3}
\end{eqnarray}
Equation~(\ref{EqMaxwell3}) has a unique non-trivial solution given by
\begin{equation}
\zeta_{H}=-\zeta_{L}=\sqrt{-\frac{\mathcal{B}_{3}}{\mathcal{B}_{5}}t}.
\label{Eqnusl}
\end{equation}
According to the numerical results for $\mathcal{B}_{3}$ and $\mathcal{B}_{5}$ in Table~\ref{qphiplane}, the argument under the square root function keeps positive, considering that the low/high-$\Phi$ BH phase transition occurs when $T>T_c$. From Eq. (\ref{Eqnusl}), one can
find that
\begin{equation}
\vartheta=\Phi_{c}(\zeta_{H}-\zeta_{L})=2\Phi_{c}\zeta_{H}=2\Phi_{c}\sqrt{-\frac{\mathcal{B}_{3}}{\mathcal{B}_{5}}t}
~~\Longrightarrow~~\chi=\frac{1}{2}.
\end{equation}

Now, we can differentiate Eq. (\ref{Eqnu2}) to calculate the
critical exponent $\sigma$ as
\begin{equation}
\kappa_{T@\langle q-\Phi\rangle}=-\frac{1}{\Phi}\frac{\partial \Phi}{\partial q_2}\bigg|_{\Phi_c}=
\frac{-1}{ \mathcal{B}_{3}  q_{c}}t^{-1}~~\Longrightarrow~~\sigma=1.
\end{equation}

Finally, the shape of $|q_2-q_c|$ at the critical isotherm $t = 0$ is given by
\begin{equation}
|q_2-q_c|=q_c|\xi-1|=q_c\left|\mathcal{B}_5\zeta^3\right|=\frac{q_c\left|\mathcal{B}_5\right|}{\Phi_c^3}|\Phi-\Phi_c|^3,
\end{equation}
thus we have $\iota=3$.

The obtained results show that the critical exponents in this new approach (with fixed $\Lambda $ and variable $q$) are the same as those obtained in the former section (with variable $\Lambda$ and fixed $q$) and coincide with the van der Waals fluid system~\cite{Kubiznak:2012wp}. For fixed dimensionality and range of interactions, the critical exponents are independent of the details of a physical system, and therefore, one may regard them quasi-universal.

\section{Phase structures of AdS BHs surrounded by CDF using shadow analysis}
\label{section5}

The images of the supermassive BHs in the galaxies $M87^*$~\cite{EventHorizonTelescope:2019dse} and $Sgr A^*$~\cite{EventHorizonTelescope:2022xnr} given by the event horizon telescope (EHT), displaying a dark part surrounded by a bright ring, are direct supports of the existence of the BH in the Universe. If light
passes close to a BH, the rays can be deflected very strongly and even travel on circular orbits. This strong deflection, together with the fact that no light comes out of a BH, has the effect that a BH is seen as a dark disk in the sky; this disk is known as the BH shadow. The BH shadow is a useful tool to understand the fundamental properties of the BH and reveal physical constrains on the gravitational theories. Recently, people are interested in studying the relation between the shadow and the thermodynamical phase transition for AdS BH. Zhang and Guo~\cite{Zhang:2019glo} found that the phase structure can be reflected by the shadow radius for the spherically symmetric BH, and that the shadow size gives correct information but the distortion of the shadow gives wrong information of the phase structure for the axially symmetric BH. Belhaj and Moumni etc.~\cite{Belhaj:2020nqy} studied the relations between the BH shadow and charged AdS BH critical behavior in the extended phase space. Hendi and Jafarzade~\cite{Hendi:2020ebh} investigated the relations between shadow radius and phase transitions for charged quintessence-surrounded AdS BH, and calculated the critical shadow radius where the BH undergoes a second-order phase transition. Guo and Li etc.~\cite{Guo:2022yjc} structured the dependence of the regular Bardeen-AdS BH shadow and thermodynamics. Du and Li etc.~\cite{Du:2022hom} investigated the relationship between the shadow radius and the first-order phase transition for the non-linear charged AdS BH in the frame of the Einstein-power-Yang-Mills gravity.

In this section, we dedicate to investigate the relation between shadow radius and phase transitions for the AdS BHs surrounded by CDF. We employ the Hamilton-Jacobi method for a photon in the BH spacetime. The Hamilton-Jacobi equation is expressed as~\cite{Decanini:2011xw}
\begin{equation}
\frac{\partial S}{\partial \sigma}+H=0,
\end{equation}
where $S$ and $\sigma$ are the Jacobi action and affine parameter along
the geodesics, respectively. The Hamiltonian of the photon moving in the
static spherically symmetric spacetime is
\begin{equation}
H= \frac{1}{2}g^{\mu\nu}\frac{\partial S}{\partial x^{\mu}}\frac{\partial S}{%
\partial x^{\nu}}=0.  \label{EqHamiltonian}
\end{equation}

Due to the spherically symmetric property of the BH, one can
consider a photon motion on the equatorial plane with $\theta=\frac{\pi}{2}$. So, Eq. (\ref{EqHamiltonian}) reduces to
\begin{equation}
\frac{1}{2}\left[-\frac{1}{f(r)}\left(\frac{\partial H}{\partial\dot{t}}
\right)^{2}+f(r)\left(\frac{\partial H}{\partial\dot{r}} \right)^{2}+\frac{1%
}{r^{2}}\left(\frac{\partial H}{\partial\dot{\phi}} \right)^{2} \right]=0.
\label{EqNHa}
\end{equation}

Regarding the fact that the Hamiltonian does not depend explicitly
on the coordinates $t$ and $\phi$, one can define
\begin{equation}
\frac{\partial H}{\partial\dot{t}}=-E ~~~~and~~~~ \frac{\partial H}{\partial%
\dot{\phi}}=L ,  \label{Eqenergy}
\end{equation}
where constants $E$ and $L$ are, respectively, the energy and angular
momentum of the photon. Using the Hamiltonian formalism, the equations of
motion are obtained as
\begin{eqnarray}
\dot{t} &=&\frac{dt}{d\sigma}=-\frac{1}{f(r)}\left(\frac{\partial H}{\partial%
\dot{t}} \right)~~~, \\
&&  \nonumber \\
\dot{r}&=&\frac{dr}{d\sigma}=-f(r)\left(\frac{\partial H%
}{\partial\dot{r}} \right), \\
&&  \nonumber \\
\dot{\phi}&=&\frac{d\phi}{d\sigma}=\frac{1}{%
r^{2}}\left(\frac{\partial H}{\partial\dot{\phi}} \right).
\label{eqEOM}
\end{eqnarray}

By using the radial equation of motion
\begin{equation}
\dot{r}^{2}+\mathcal{V}_{eff}(r)=0,  \label{radialEoM}
\end{equation}
the effective potential of the photon can be obtained as
\begin{equation}
\mathcal{V}_{eff}(r)=f(r)\left[ \frac{L^{2}}{r^{2}}-\frac{E^{2}}{f(r)}\right].  \label{Eqpotential}
\end{equation}

\begin{figure}[htb]
\includegraphics[width=0.9\linewidth]{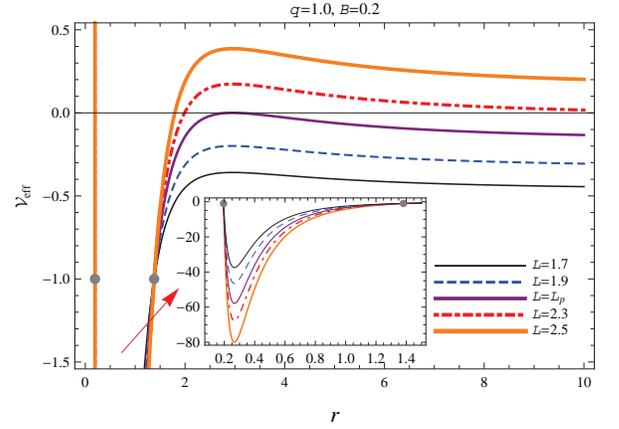}
\caption{\label{FigVef} Effective potential $\mathcal{V}_{eff}$ as a function of $r$ for $E=1.0$, $M=1.0$, $\Lambda=-1$ and various $L$. Here, the critical angular momentum $L_p=2.12318$. The gray points are related to the positions of the inner and outer horizons, where $\mathcal{V}_{eff}=-1.0$.}
\end{figure}
Fig. \ref{FigVef} depicts the behavior of the photon's effective potential for $E =1$ with various $L$. As we see, there exists a peak of the effective potential which increases with increasing $L$. Due to the constraint $ \dot{r}^{2} \geq 0 $, we expect that the effective potential satisfies $ \mathcal{V}_{eff}\leq 0 $. So, an ingoing photon from infinity with the negative effective potential falls into the BH inevitably, whereas it bounces back if $\mathcal{V}_{eff}> 0 $. For small $L$, the photon can fall into the BH from a place with large $r$. However, for large $L$, the peak of the potential will above zero, then the photon will be reflected before it falls into the BH. Between the two cases, there exists a critical case described by the purple solid line (with medium thickness), $L = L_{p}$ ($ \mathcal{V}_{max,eff} = 0$). At its peak point $r = r_{max,L_{p}}$, the photon has zero radial velocity and acceleration. So the photon will round the BH at that radial distance. For a static spherically symmetric BH, it corresponds to the photon sphere with radius $r_{p}=r_{max,L_{p}}$. From what was expressed, one can find that the photon orbits are circular and unstable associated to the maximum value of the effective potential. In order to obtain such a maximum value, we use the following conditions, simultaneously
\begin{equation}
\mathcal{V}_{eff}(r)\Bigg\vert_{r=r_{p}}=0,~~~~~\frac{\partial \mathcal{V}_{eff}(r)}{\partial r}%
\Bigg\vert_{r=r_{p}}=0,
\label{Eqcondition}
\end{equation}
determining the critical angular
momentum of the photon sphere ($ L_{p} $) and the photon sphere
radius ($ r_{p} $), respectively, resulting in the following equation
\begin{equation}
f\left(r_p\right)\left(6M-2r_p-q{\rm ArcSinh}\frac{q}{\sqrt{B}r_p^3}\right)r_p=0.
\label{Eqrpn0}
\end{equation}
Since the photon sphere radius should be larger than the event horizon radius for a BH, the photon sphere radius satisfies $f\left(r_p\right)>0$. Then Eq.~(\ref{Eqrpn0}) leads to
\begin{equation}
6M-2r_p-q{\rm ArcSinh}\frac{q}{\sqrt{B}r_p^3}=0.
\label{Eqrpn1}
\end{equation}
It follows that the photon sphere radius does not depend on the cosmological constant. For the circular orbit of the photon, we also have the following constraint
\begin{equation}
\frac{\partial^{2} \mathcal{V}_{eff}(r)}{\partial r^{2}}\Bigg\vert_{r=r_{p}}<0,
\label{Eqcondition2}
\end{equation}
to ensure that the photon orbits are unstable.

The orbit equation for the photon is obtained in the
following form
\begin{equation}
\frac{dr}{d\phi}=\frac{\dot{r}}{\dot{\phi}}=\frac{r^{2}f(r)}{L}\left(\frac{%
\partial H}{\partial\dot{r}} \right).  \label{Eqorbit}
\end{equation}

The turning point of the photon orbit is expressed by the following
constraint
\begin{equation}
\frac{dr}{d\phi}\Bigg\vert_{r=R}=0.  \label{EqTpoint}
\end{equation}

Using Eqs. (\ref{EqNHa}) and (\ref{EqTpoint}), one gets
\begin{equation}
\frac{dr}{d\phi}=\pm r\sqrt{f(r)\left[\frac{r^{2}f(R)}{R^{2}f(r)} -1\right] }%
.  \label{EqNorbit}
\end{equation}

Considering a light ray sending from a static observer placed at $r_{o} $
and transmitting into the past with an angle $\alpha$ with respect to the
radial direction, one can write~\cite{Zhang:2019glo}
\begin{equation}
\cot \alpha =\frac{\sqrt{g_{rr}}}{g_{\phi\phi}}\frac{dr}{d\phi}\Bigg\vert%
_{r=r_{o}}.  \label{Eqangle}
\end{equation}

Hence, the shadow radius of the BH can be obtained as
\begin{equation}
r_{s}=r_{o}\tan \alpha\approx r_{o}\sin \alpha =R\sqrt{\frac{f(r_{o})}{f(R)}}\Bigg\vert_{R\rightarrow r_{p}},
\label{Eqshadow}
\end{equation}
where $r_{o}$ is the position of the observer. We mention that the approximation implemented in Eq.~(\ref{Eqshadow}) is valid only for small value of $\alpha$.
\begin{figure}[htb]
\includegraphics[width=0.9\linewidth]{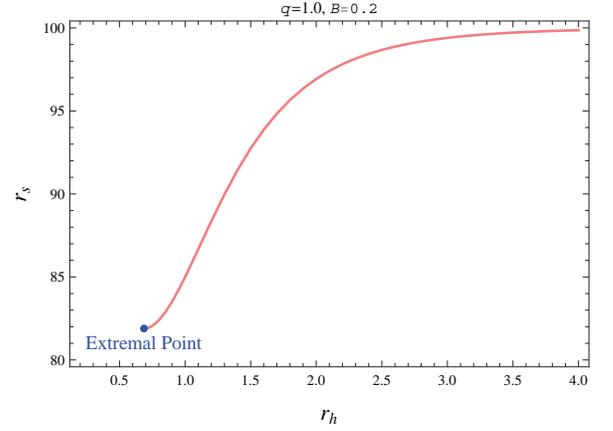}
\caption{\label{rhrs} The variation of shadow radius $r_s$ in terms of the event horizon radius $r_h$, here we have set $\Lambda=-1$ and $r_o=100$. The extremal point corresponds to the horizon radius $r_{ext.}=0.68666$, at which the inner and outer horizons coincide.}
\end{figure}
\begin{figure*}[htb]
\begin{tabular}{ c c }
\includegraphics[width=0.430\linewidth]{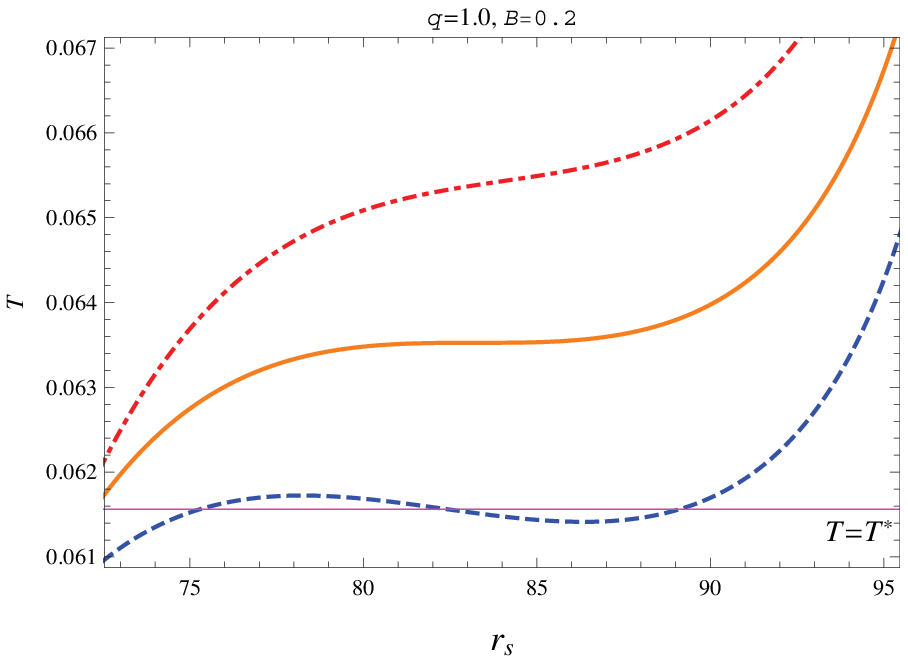}
\hspace{.10in}
\includegraphics[width=0.435\linewidth]{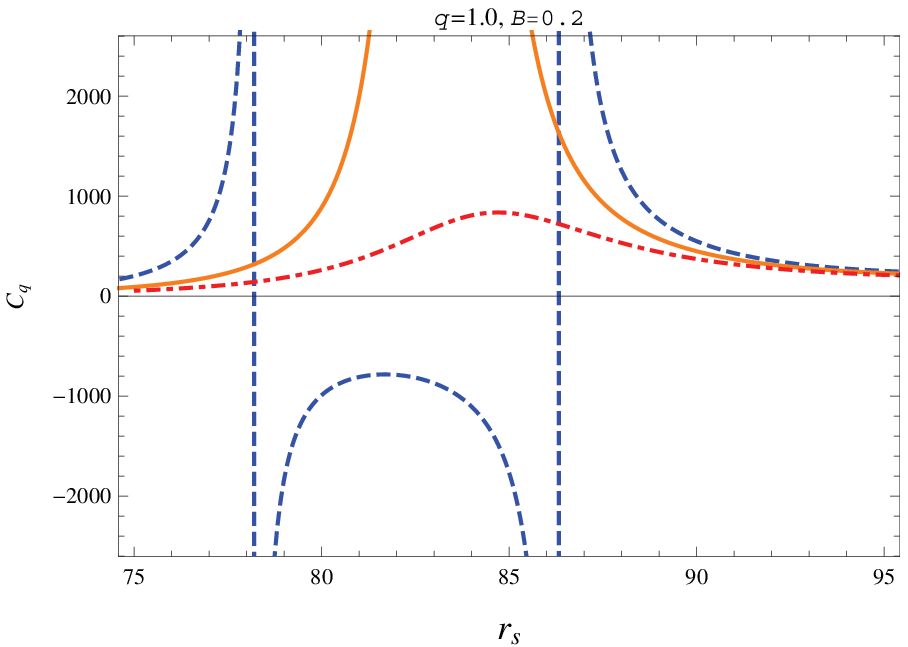}
\end{tabular}
\caption{\label{TCqrs} Hawking temperature (left panel) and the heat capacity (right panel) with respect to the BH shadow radius $r_s$ for an observer at $r_o=100$. The blue (dashed), orange (solid) and red (dot-dashed) curves correspond to the curves with $P=0.98P_c$, $P=1.00P_c$, and $P=1.02P_c$, where $P_c=0.02441$. Also, the blue (dashed) line corresponds to the location of the first-order phase transition, while the orange (solid) line to the second-order one. $T^*$ in the $T-r_s$ panel represents the coexistence temperature.}
\end{figure*}
Since Eqs. (\ref{Eqrpn1}) and (\ref{Eqshadow}) are complicated to solve analytically, we employ numerical methods to obtain the radius of the photon sphere and shadow. With the help of Eq.~(\ref{BHMass}), the
behaviors of the shadow radius as the function of the BH horizon radius can be numerically obtained, which are exhibited in Fig.~\ref{rhrs}. We observe that there exists a positive correlation between the shadow radius and the event horizon radius, indicating that the shadow radius could be a fine quantity reflecting the phase structure of the static spherically symmetric AdS BH surrounded by CDF.
\begin{figure}[htb]
\includegraphics[width=0.9\linewidth]{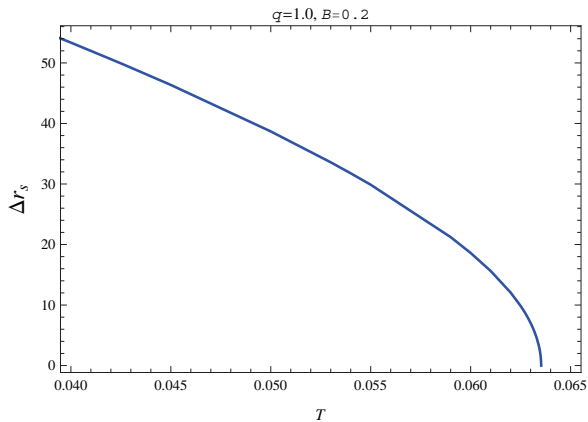}
\caption{\label{Tdeltars} The $\Delta r_s$ as function of the coexistence temperature $T$, here we have set $r_o=100$.}
\end{figure}
Now, we are interested in examining the relations between the
shadow radius and phase transitions. According to
\cite{Zhang:2019glo,Belhaj:2020nqy,Guo:2022yjc,Du:2022hom,Wei:2017mwc,Li:2019dai}, there is a close connection
between BH shadows and the BH thermodynamics. The heat capacity is
one of the interesting thermodynamic quantities which provides the
information related to the thermal stability and phase transition
of a thermodynamic system. The sign of heat capacity determines
thermal stability/instability of BHs. The positivity
(negativity) of this quantity indicates a BH is thermally stable
(unstable). Besides, the discontinuities in heat capacity could be
interpreted as the possible phase transition points. According to Eq. (\ref{Cq}), heat capacity can be written as
\begin{equation}
C_q=T\left(\frac{\partial S}{\partial r_{h}}\frac{\partial
r_{h}}{\partial T} \right)_q.
\end{equation}
By using the fact that $ \frac{\partial S}{\partial r_{h}}>0 $,
the sign of $ C_q$ is directly inducted from $ \frac{\partial
T}{\partial r_{h}} $ which can be rewritten as
\begin{equation}
\frac{\partial T}{\partial r_{h}}=\frac{\partial T}{\partial r_{s}}\frac{\partial r_{s}}{\partial r_{h}}.
\end{equation}

Since the shadow radius is positively correlated with the event horizon radius, i.e. $\frac{\partial r_{s}}{\partial r_{h}}>0$, one can draw a conclusion that the sign of $ C_q$ is controlled by $ \frac{\partial T}{\partial r_{s}} $, which leads us to the study on the behaviors of temperature and heat capacity with respect to $r_{s}$ by using numerical technique. The isobar curves on the $T-r_{s}$ and $C_q-r_{s}$ panels are displayed in Fig.~\ref{TCqrs}. As we see, $C_q-r_{s}$ curves exhibit similar behaviors as $C_q-r_h$ curves (shown in the left panel of Fig.~\ref{Cvphi}) for $P>P_c$, $P<P_c$ and $P=P_c$. For $P>P_{c}$, the temperature is only a monotone increasing function of $r_{s}$ without any extremum (see dot-dashed line in left panel of Fig.~\ref{TCqrs}). The heat capacity is also a continuous function for variable $r_{s}$ (see dot-dashed line in right panel of Fig.~\ref{TCqrs}). For the case $P<P_{c}$, a non-monotonic behavior appears for temperature with one local maximum and one minimum (see dashed line in left panel of Fig.~\ref{TCqrs}), which indicates the existence of first-order phase transition. According to the definition of heat capacity, these extrema points of $T$ coincide with divergence points (see dashed line in right panel of Fig.~\ref{TCqrs}) of $C_q$. Evidently, a change of signature occurs at these points. To be more specific, it changes from positive to negative at the first divergency and then it becomes positive again at the second one. So, three branches of BHs are thermodynamically competing. The BHs with shadow smaller than the first divergency of heat capacity are stable; The region after the second divergency of heat capacity is related to BHs which are also thermally stable; For an intermediate range of shadow, the BHs are thermodynamically unstable.

One should note that, the intermediate shadow domain between the two divergence points in specific heat diagram does not coincide exactly with the coexistence area of the small/large BH phase transition. To describe the phase transition more precisely in $T-r_s$ plane, one has to replace the `oscillating part' of the isobar by an isotherm. To accomplish this, one should first acquire the coexistence temperature $T^*$ by using the Maxwell equal area law in $P-V$ plane, then determine the horizon radii by equating $T(r_h^{s,l})=T^*$, and finally calculate the shadow radii $r_s^{s}/r_s^{l}$ for small/large BHs by utilizing Eqs.~(\ref{Eqrpn1}) and ~(\ref{Eqshadow}). We depict the changes of the shadow radius ($\Delta r_{s}=r_{s}^{l}- r_{s}^{s}$) as a function of coexistence temperature in Fig.~\ref{Tdeltars}. We see that $\Delta r_{s}$ has similar behavior with $\Delta V$ (shown in the right panel of Fig.~\ref{Pvphasestructrue}) and it is monotone decreasing function of $T$. At $P=P_{c}$, the small BH and the large BH merge into one squeezing out the unstable BH. $\Delta r_{s}$ approaches to zero at $T=T_{c}$, which is exactly the coexistence temperature at $P=P_c$ for the $T-r_s$ relation in Fig.~\ref{TCqrs}, where the first-order phase transition becomes a second-order one. Such behavior of temperature is very similar to van der Waals liquid/gas system which undergoes a second-order phase transition at $T=T_{c}$. Through the above analysis, we conclude that we can detect whether there is a phase transition by measuring the shadow radius of the BH. If one observes a sudden change of $r_s$, then the BH system must experience a first-order phase transition; If a $T-r_{s}$ curve deflection point is observed, then the BH system experiences a second-order phase transition.

\section{Conclusion}
\label{section6}
Dark fluids, including dark matter and dark energy, contribute most to the ingredients of the Universe. A cosmological dark fluid with Chaplygin-like equation of state $p=-\frac{B}{\rho}$ could be a naturally existing substance, considering its amusing connection with string theory. Motivated by this, we have derived an exact static spherically-symmetric AdS BH solution endowed with CDF background. The energy density, radial and tangential pressures of the CDF have also been calculated. We have examined the classical energy conditions for the CDF and found that it admits the null, weak and dominant energy conditions, while violates the strong energy condition. The first law of thermodynamics, as well as the Smarr relation have been constructed. Interestingly, the thermodynamic quantities, consisting of the mass (interpreted as enthalpy), temperature, specific heat and the Gibbs free energy are corrected, while the thermodynamic volume and entropy are not directly affected by the CDF.

In turn, we have investigated the extended phase space of thermodynamics and studied the critical phenomena of the AdS BHs surrounded by CDF by treating the cosmological constant as a thermodynamic pressure. We have found a first-order small/large BH phase transition, which is analogous to the liquid/gas phase transition in van der Waals fluid. The critical exponents coincide with those of the Van der Waals fluid. We have found that, $r_c$, $P_c$ and $T_c$ depend differently on $q$ and $B$, and the ratio $\frac{P_cr_c}{T_c}$ does not keep constant as the charged AdS BH does, this phenomenon reflects again the effects of the CDF.

In the non-extended phase space with variable value of normalization factor $q$ and fixed cosmological constant $\Lambda$, to deduce the equation of state of the BH, $q$ has been solved as a double-valued function of $\Phi$ and $T$. It has been found that such a BH admits a van der Waals-like first order low/high-$\Phi$ BH phase transition and possesses the same critical exponents with van der Waals fluids. Also the effects of the parameters $B$ and $P$ on critical quantities, $\Phi_c$, $q_c$ and $T_c$, have been numerically studied.

Finally, we have studied the shadow thermodynamics of AdS BH surrounded by CDF. It has been found that, the shadow radius and the event horizon radius display a positive correlation. By analyzing the phase transition curves under the shadow context, we have found that the shadow radius can replace the event horizon radius to present the BH phase transition process, and the phase transition grade can also be revealed by the shadow radius, indicating that the shadow radius may serve as a probe for the phase structure in our case.

\appendix
\section{Expressions for $B_i$ parameters in Eq.~(\ref{EqaTc1})}
\label{appendix1}
The expressions for $B_i$ are listed in Table~\ref{Bi}.
\begin{table*}[htp]
\caption{Expressions for $B_i$ in Eq.~(\ref{EqaTc1}).}~\label{Bi}
\begin{tabular}{|c|c|}
  \hline
  $B_i$ & \textbf{Parameter Expression}\\ \hline
  $B_1$ & $\sqrt{-P+4\pi^2 T_c^2+\sqrt{B}{\rm Cosh}6\Phi_c}$\\ \hline
  $B_2$ & $P-\sqrt{B}{\rm Cosh}6\Phi_c$ \\ \hline
  $B_3$ & $-\frac{1}{2\pi}P+T_c(-B_1+2\pi T_c)$ \\ \hline
  $B_4$ & $\sqrt{B}-8\pi B_3{\rm Cosh}6\Phi_c-5\sqrt{B}{\rm Cosh}12\Phi_c+12B_1(Bq_c)^{1/3}\left({\rm Sinh}6\Phi_c\right)^{5/3}$ \\ \hline
  $B_5$ & $\sqrt{B}-8\pi B_3{\rm Cosh}6\Phi_c-5\sqrt{B}{\rm Cosh}12\Phi_c+18B_1(Bq_c)^{1/3}\left({\rm Sinh}6\Phi_c\right)^{5/3}$ \\\hline
  $B_6$ & $5\sqrt{B}-\sqrt{B}{\rm Cosh}12\Phi_c+4(-P+4\pi^2 T_c^2){\rm Cosh}6\Phi_c$ \\\hline
  $B_7$ & $B^{1/6}B_2(-B_6\pi T_c+2B_1^3{\rm Cosh}6\Phi_c)q_c^{2/3}\left({\rm Csch}6\Phi_c\right)^{2/3}+B_1B_5(B_1-2\pi T_c)B^{1/3}q_c^{1/3}\left({\rm Csch}6\Phi_c\right)^{1/3}$ \\\hline
  $B_8$ & $2B_1(B_1-2\pi T_c){\rm Cosh}6\Phi_c+3\sqrt{B}\left({\rm Sinh}6\Phi_c\right)^2$ \\\hline
  \multirow{2}{*}{\centering $B_{9}$} & $27B+80B_1^4-160\pi B_1^3T_c+60\sqrt{B}B_1^2{\rm Cosh}6\Phi_c-60\sqrt{B}B_1^2{\rm Cosh}18\Phi_c+9B{\rm Cosh}24\Phi_c$\\
                                      & $-4\left[9B+4B_1^3(B_1-2\pi T_c)\right]{\rm Cosh}12\Phi_c$\\ \hline
  $B_{10}$ & $2 P {\rm Cosh}6\Phi_c+\sqrt{B}\left(-5+3{\rm Cosh}12\Phi_c\right)$ \\\hline
  $B_{11}$ & $\frac{16B^{1/6}B_8^2}{B_1^2}\left({\rm Csch}6\Phi_c\right)^{4/3}+\frac{B_2B_9q_c^{1/3}}{B_1^3}\left({\rm Csch}6\Phi_c\right)^{5/3}+144B^{1/3}B_{10}q_c^{2/3}-\frac{288\sqrt{B}B_8q_c^{1/3}}{B_1}\left({\rm Sinh}6\Phi_c\right)^{1/3}$ \\\hline
  $B_{12}$ & $-10B+B_2^2+12\sqrt{B}B_2{\rm Cosh}6\Phi_c+10B{\rm Cosh}12\Phi_c$ \\\hline
  $B_{13}$ & $\sqrt{2}P-5\sqrt{2B}{\rm Cosh}6\Phi_c+\left(\sqrt{2}P+3\sqrt{2B}{\rm Cosh}6\Phi_c\right){\rm Cosh}12\Phi_c$ \\\hline
  $B_{14}$ & $8B^{1/6}B_1B_8^2\left({\rm Csch}6\Phi_c\right)^2+B_2B_9q_c^{1/3}\left({\rm Csch}6\Phi_c\right)^{7/3}$ \\\hline
  $B_{15}$ & $16B^{1/6}B_1B_8^2\left({\rm Csch}6\Phi_c\right)^{1/3}+B_2B_9q_c^{1/3}\left({\rm Csch}6\Phi_c\right)^{2/3}$ \\ \hline
  \multirow{3}{*}{\centering $B_{16}$} &  $2448\sqrt{B}B_1^4-810B^{3/2}-96B_1^2\left[9B-26B_1^3\left(B_1-2\pi T_c\right)\right]{\rm Cosh}6\Phi_c+27\sqrt{B}\left(45B-128B_1^4\right){\rm Cosh}12\Phi_c$\\
                        &  $+1296BB_1^2{\rm Cosh}18\Phi_c+64B_1^6{\rm Cosh}18\Phi_c-128\pi B_1^5T_c{\rm Cosh}18\Phi_c-486B^{3/2}{\rm Cosh}24\Phi_c$ \\
                        & $+1008\sqrt{B}B_1^4{\rm Cosh}24\Phi_c-432BB_1^2{\rm Cosh}30\Phi_c+81B^{3/2}{\rm Cosh}36\Phi_c$\\ \hline
  $B_{17}$ & $-\frac{8B_1B_8B_{14}}{B_2}\left(\frac{B}{q_c}\right)^{1/3}-B_2B_{16}B^{1/6}q_c^{1/3}\left({\rm Csch}6\Phi_c\right)^{8/3}-16B^{1/3}B_1B_8B_9\left({\rm Csch}6\Phi_c\right)^{7/3}$ \\\hline
  $B_{18}$ & $216B^{2/3}B_1^2B_{15}+B_2B_{17}-2592\sqrt{2}B_8B_{13}B_1^4q_c^{1/3}B^{2/3}\left({\rm Sech}6\Phi_c\right)^{1/3}+6912\sqrt{B}B_1^5B_{12}q_c^{2/3}{\rm Sinh}6\Phi_c$ \\\hline
\end{tabular}
\end{table*}
\acknowledgements
This work is supported by the Special Foundation for Theoretical Physics Research Program of China (Grant No. 11847065), the Natural Science Foundation of Shanxi Province (No. 201901D211110). The authors would like to express their gratitude for the detailed discussions provided by the anonymous reviewer, which greatly contributed to improving the quality of the paper.


\end{document}